\documentclass{llncs}

\usepackage[american]{babel}
\usepackage{latexsym}
\usepackage{amssymb} 
\usepackage{amsmath}
\usepackage{amsbsy}
\usepackage{cite}
\usepackage{framed,xspace}
\usepackage{url}
\usepackage{qip}
\usepackage[active]{srcltx}
\usepackage{afterpage}
\usepackage{rotating}

\newenvironment{proofx}[1][] {\noindent{\bf Proof#1.}\hspace{0.75em}}
	       {\hspace{\fill}$\blacksquare$\vspace{0.3cm}}

\newcommand{\dP}{{\sf P}'}

\newcommand{\A}{{\sf A}}
\newcommand{\dA}{{\sf A}'}
\newcommand{\hA}{\hat{\sf A}}
\newcommand{\dhA}{\hat{\sf A}'}

\newcommand{\B}{{\sf B}}
\newcommand{\dB}{{\sf B}'}

\newcommand{\hB}{\hat{\sf B}}
\newcommand{\dhB}{\hat{\sf B}'}

\newcommand{\dPlayer}{\mathfrak{P}}
\newcommand{\dPlayerPoly}{\dPlayer}

\newcommand{\negl}[1]{{\sc{negl}}\,(#1)}

\newcommand{\approxs}{\stackrel{\text{\it\tiny s}}{\approx}}    
\newcommand{\approxq}{\stackrel{\text{\it\tiny q}}{\approx}}    

\newcommand*{\assign}
	   {\ensuremath{\kern.5ex\raisebox{.1ex}{\mbox{\rm:}}\kern
	   -.3em =}}

\newcommand{\set}[1]{\{#1\}}

\newcommand{\zo}{\{0,1\}}

\newcommand{\X}{\mathcal{X}}

\newcommand{\F}{\mathcal{F}} 

\newcommand{\commitk}[3]{\mathtt{commit}\,_{#3}\,(#1,#2)\,}
\newcommand{\commit}{\mathtt{commit}}

\newcommand{\pk}{{\tt pk}}

\newcommand{\GH}{{\cal G}_{\tt H}}
\newcommand{\GB}{{\cal G}_{\tt B}}

\newcommand{\MC}[3]{#1 \leftrightarrow #2 \leftrightarrow #3}

\newcommand{\prob}[1]{\operatorname{Pr \,}[#1]}
\newcommand{\la}{\leftarrow}
\newcommand{\ra}{\rightarrow}

\newcommand{\COIN}{\mathtt{COIN}}
\newcommand{\cxPi}[3]{\Pi_{#1,#2}^{\, #3-\COIN}}  

\newcommand{\cxF}[1]{\F_{{#1}-\COIN}}	
\newcommand{\clF}{\F_{\ell-\COIN}}	

\newcommand{\uncont}{\texttt{uncont}}
\newcommand{\force}{\texttt{force}}
\newcommand{\random}{\texttt{random}}
\newcommand{\defterm}[1]{\emph{#1}}

\newcommand{\xtr}[2]{{\tt{xtr}}_{#2}(#1)}
\newcommand{\xtrx}{\tt{xtr}}

\newcommand{\sss}{\mathtt{sss}}

\newcommand{\Commit}{\mathtt{COMMIT}}
\newcommand{\Commitk}[3]{\mathtt{COMMIT}\,_{#3}\,\big( #1,#2 \big)\,}

\newcommand{\FF}{\mathbb{F}}
\newcommand{\NN}{\mathbb{N}}

\newcommand{\D}[1]{\mathcal{D}^{\, #1}}
\newcommand{\U}{\mathcal{U}}

\newcommand{\start}{\mathtt{start}}

\newcommand{\abort}{\texttt{abort}}
\newcommand{\success}{\texttt{success}}

\newcommand{\ZKPK}{\mathtt{ZKPK(\Rel)}}

\newcommand{\zkpkF}{\F_{\ZKPK}}
\newcommand{\Rel}{\mathcal{R}}
\newcommand{\NP}{\ensuremath{\mathcal{NP}}}

\newcommand{\OT}{\mathtt{OT}}
\newcommand{\sfeF}{\F_{\SFE}^f}
\newcommand{\SFE}{\mathtt{SFE}}
\newcommand{\Sim}{\mathcal{S}}

\newcommand{\emphbf}[1]{\emph{\textbf{#1}}}

\begin{document}

\pagestyle{plain}

\title{Fully Simulatable Quantum-Secure Coin-Flipping and
  Applications} 
\author{Carolin Lunemann \and Jesper Buus Nielsen}
\institute{Department of Computer Science, Aarhus University,
  Denmark\\ \email{\{carolin|jbn\}@cs.au.dk} }

\maketitle


\begin{abstract}
  We propose a coin-flip protocol which yields a string of strong,
  random coins and is fully simulatable against poly-sized quantum
  adversaries on both sides. It can be implemented with
  quantum-computational security without any set-up assumptions, since
  our construction only assumes mixed commitment schemes which we show
  how to construct in the given setting. We then show that the
  interactive generation of random coins at the beginning or during
  outer protocols allows for quantum-secure realizations of classical
  schemes, again without any set-up assumptions. As example
  applications we discuss quantum zero-knowledge proofs of knowledge
  and quantum-secure two-party function evaluation. Both applications
  assume only fully simulatable coin-flipping and mixed
  commitments. Since our framework allows to construct fully
  simulatable coin-flipping from mixed commitments, this in particular
  shows that mixed commitments are \emph{complete} for quantum-secure
  two-party function evaluation. This seems to be the first
  completeness result for quantum-secure two-party function evaluation
  from a generic assumption.
\end{abstract}


\section{Introduction}
\label{sec:intro}

True randomness is a crucial ingredient in many cryptographic
applications. Therefore, secure coin-flipping is an essential
primitive, which allows two parties to agree on a uniformly random bit
in a fair way, such that neither party can influence the value of the
coin to his advantage. We investigate coin-flip protocols with
classical messages exchange but where the adversary is assumed to be
capable of quantum computing. Security of cryptographic protocols in
the quantum world means, of course, that quantum computation does not
jeopardize the assumption, underlying the protocol
construction. However, we encounter additional setbacks in the
security proofs, which are mostly due to the fact that some well-known
classical proof techniques cannot be applied in a quantum environment.

\noindent{\sc Our Contribution.}  We aim at establishing coin-flipping
as a stand-alone tool in a model without any setup assumptions. As
such, our protocol can be used in several contexts and different
generic constructions. One notable application is as subroutine for
realizing the theoretical assumption of the common-random-string-model
(CRS-model).\footnote{In the CRS-model the parties are provided with a
  public common random string CRS before communication, taken from the
  uniform distribution.} Since the generation of a CRS often
significantly simplifies the design of (quantum-secure) protocols,
this then implies that various interesting applications can be
implemented quantum-securely in a simple manner from scratch.

In more detail, we first investigate different degrees of security
that a coin-flip protocol can acquire. Then, we propose and prove
constructions that allow us to amplify the respective degrees of
security such that weaker coins are converted into very strong
ones.\footnote{For clarity, we note that we use the intuitive
  interpretation of ``weak'' and ``strong'' coins related to their
  security degrees, which differs from the definitions in the quantum
  literature.} The amplification only requires mixed commitment
schemes, which we know how to construct with quantum security under
reasonable assumptions---for instance, based on the quantum hardness
of the learning with error problem. Combining our amplification
protocols allows to take a very weak notion of coin-flipping and
amplify it to a coin-flip protocol which is \defterm{fully simulatable
  against poly-sized quantum adversaries}. By fully simulatable we
mean that both sides can be simulated in quantum polynomial time.

Our amplification framework should also be understood as a step
towards fully simulatable \emph{constant-round} coin-flipping. To the
best of our knowledge, to date there does not exist any fully
simulatable protocol which is constant-round and which allows to
generate a long random bit-string. In particular, no fully simulatable
constant-round coin-flip protocol is known to securely compose in
parallel. Since all our amplification protocols work in
constant-round, we show that if there exists a constant-round
coin-flip protocol of long strings with weak security, then there also
exists a constant-round coin-flip protocol of long strings which is
fully simulatable. Even though our work leaves fully simulatable
constant-round coin-flipping of long strings as a fascinating open
problem, we consider it a contribution in itself to define a
reasonably weak but sufficient security notion to realize fully
simulatable constant-round coin-flipping of long strings.

\noindent{\sc Related Work.} The standard coin-flip protocol
of~\cite{Blum81} was proven secure in a quantum environment in
previous work~\cite{DL09}. In its basic form this protocol yields
\emph{one} coin as output. Of greater importance, however, is flipping
a string of coins instead of a bit, in particular, when generating a
CRS. The basic construction composes in sequence with security
classified as medium in our framework here. Parallel composition is
possible using an extended construction providing efficient
simulations on both sides. This extension, however, requires a CRS as
initial assumption, i.e.\ the CRS-model, and hence, violates our
strong requirement of applications, implementable quantum-securely
without any set-up assumptions.

As an example application, we discussed in~\cite{DL09} the generation
of a CRS in the context of e.g.\ a quantum zero-knowledge proof. For
an overview and more details, see also~\cite{Lunemann10}. To further
show the implications of coin-flipping as an implementation of the
CRS-model in the quantum setting, we here add the functionalities of a
quantum zero-knowledge proof of knowledge and quantum-secure function
evaluation. We want to mention the following related work. First, an
alternative approach in the context of zero-knowledge was
independently investigated by Smith~\cite{Smith09}. There,
coin-flipping is implemented by a string commitment with special
openings and validated in subsequent zero-knowledge proofs in
sequence, and which therefore has round complexity depending on the
security parameter, i.e.\ how many proofs must be completed to achieve
a negligible soundness error. The coin-string is used as key to encode
the witness and more zero-knowledge proofs are given to prove that. As
encryption scheme, they suggest a scheme with similar properties as in
the standard construction for mixed
commitments~\cite{DN02,DFS04,DFLSS09}. To the best of our knowledge,
the question of its actual secure implementation was left open, and a
formal description and analysis was never published. Second, we want
to mention the concurrent and independent work of Hallgren, Smith, and
Song, as sketched in~\cite{HSS11}. They also prove, among other
things, classical protocols for zero-knowledge proofs of knowledge and
function evaluation secure in the quantum setting by proposing a
composition theorem that allows to use the basic coin-flipping
protocol in~\cite{DL09} to generate a CRS. In addition, they give a
UC-secure protocol for said tasks in the CRS-model.

Furthermore, the techniques used in our reductions are inspired by
techniques used by works in the UC framework (cf.~\cite{DN02}), where
rewinding is also a problem. But to the best of our knowledge, all our
reductions are novel, and might be also of classical interest.

\noindent{\sc Security in the Quantum World.} It is well known that
bit commitments imply a single coin-flip---in the classical as in the
quantum world~\cite{Blum81,DL09}---in a straightforward way: Alice
chooses a random bit $a$ and commits to it, Bob then sends his bit $b$
in plain, then the commitment is opened, and the resulting coin is $a
\oplus b$. However, even when basing the embedded commitment scheme on
a computational assumption that withstands quantum attacks (for the
hiding property), the security proof of the outer coin-flipping (and
its integration into other applications) cannot easily be translated
from the classical to the quantum world. Typically, security against a
classical adversary is argued in this context by rewinding the
adversary in a simulation. In brief, it is shown that a run of a
protocol between a dishonest Bob and honest Alice can be efficiently
simulated without interacting with Alice but with a simulator
instead. A simulator basically prepares a valid conversation and tries
it on dishonest Bob. Now, in case Bob does not send the expected
reply, we need the possibility to rewind him.  Then to conclude the
proof, we have to show that the expected running time of the
simulation is polynomial.

Unfortunately, rewinding as a proof technique can generally not be
directly applied in the quantum world, i.e., if the dishonest machine
is a quantum computer. First, we cannot trivially copy and store an
intermediate state of a quantum system, and second, quantum
measurements are in general irreversible. In order to produce a
classical transcript, the simulator would have to partially measure
the quantum system without copying it beforehand, but then it would
become impossible to reconstruct all information necessary for correct
rewinding~\cite{vdGraaf97}. It is worth mentioning though that
rewinding in the quantum world is possible in a limited setting, as
shown by Watrous~\cite{Watrous09}. This technique was also used for
proving the quantum security of single coin-flipping based on bit
commitments~\cite{DL09}. However, the generation of a string of coin
must be based on string commitments. In this setting, the simulator
cannot rewind in poly-time. A possible solutions for simulating
against a classical Bob is then to let him commit to his message in a
way which allows to extract the message in the simulation. Therewith,
the message is known to the simulator in any following iteration of
rewinding. This technique seems to be doomed to fail in the quantum
realm, since it is neither known how to rewind quantumly for string
commitments nor can any intermediate status (such as Bob's commitment)
be preserved. Moreover, commitment constructions providing flavors of
extractability without rewinding require some stronger set-up
assumptions. Thus, other techniques such as our method based on mixed
commitments, are needed for solutions in this context.

\noindent{\sc Applications.} Even though we establish coin-flipping as
a stand-alone tool, we highlight again that the generation of a CRS
leads to a simple and quantum-secure implementation of various
interesting applications without any set-up assumptions. We show two
different example applications, in addition to the functionalities
already discussed in~\cite{DL09}. First, we propose a
\emph{quantum-secure zero-knowledge proof of knowledge} based on a
witness encoding scheme, which we define such that it provides a
certain degree of extractability and simulatability in the quantum
world. Our zero-knowledge construction only requires mixed
commitments, which can be implemented with quantum security. This is
of particular interest, as the problems of rewinding in the quantum
realm complicate implementing proofs of knowledge from scratch. And
second, we show that mixed commitment schemes are sufficient for
\emph{quantum-secure function evaluation} of \emph{any} classical
poly-time function $f$ with security against active quantum
adversaries. In more detail, we first show that mixed commitments
imply an oblivious transfer protocol with passive security. From that
it is straightforward to construct a protocol for any classical
poly-time function with security against passive quantum
adversaries~\cite{Kilian88}. As our main result in that context, we
then propose a quantum-secure implementation for evaluating any such
function with security against active quantum adversaries.


\section{Preliminaries}

  \noindent{\sc Notation.} We use $\negl{n}$ to denote the set of
  \emph{negligible} functions (in $n$). For a bit-string $x \in
  \{0,1\}^n$ and a subset $S \subseteq \{ 1, \ldots, n \}$ of size
  $s$, we define $x|_S \in \{ 0,1 \}^s$ to be the restriction
  $(x_i)_{i \in S}$. The \emph{probability} of event $E$ is denoted by
  $\prob{E}$. For a random variable $X$ we use $P_X$ to denote the
  \emph{distribution} of $X$, and for an additional random variable
  $Y$ we use $P_{X\vert Y}$ to denote the \emph{conditional
    distribution} of $X$ given $Y$. \emph{Statistical
    indistinguishability} of families of classical random variables is
  denoted by $\approxs$, and $\approxq$ indicates \emph{quantum
    poly-time indistinguishability} of families of random variables,
  i.e., the families cannot be distinguished by poly-sized families of
  quantum circuits.\\

  \noindent{\sc Definition of Security.} We are interested in
  classical two-party protocols secure in a quantum world. We work in
  the security framework, introduced in~\cite{FS09} and extended
  in~\cite{DFLSS09}. The definitions are proposed for quantum
  protocols that implement \emph{classical non-reactive two-party
    functionalities}, meaning that in- and output must be
  classical. The framework allows functionalities which behave
  differently in case of a dishonest player, and it is further shown
  that any protocol in the framework \emph{composes sequentially} in a
  classical environment, i.e.\ within an outer classical protocol. For
  the sake of simplicity, the framework does not assume additional
  entities such as e.g.\ an environment. The original security
  definitions for unconditional security~\cite{FS09} are phrased in
  simple information-theoretic conditions, depending on the
  functionality, which implies strong simulation-based
  security. In~\cite{DFLSS09}, it is then shown that computational
  security (in the CRS-model) can be defined similarly. In the
  following, we state the formalism essential for this
  work.\footnote{Note that we use a simplified joint output
    representation in comparison to~\cite{FS09}.} For more details on
  the framework and notation, we refer to~\cite{FS09,DFLSS09,DFSS07},
  and to~\cite{Lunemann10} for an overview.

  Our protocols run between players Alice ($\A$) and Bob ($\B$) and
  all definitions are given in the \emph{two-world paradigm} of
  simulation-based proofs. The \emph{real world} captures the actual
  protocol $\Pi$, consisting of message exchange between the parties
  and local computations. Real-world players are denoted by honest
  $\A, \B$ and are restricted to poly-time classical
  strategies. Dishonest players $\dA, \dB$ are allowed any
  \emph{quantum} poly-time strategy. Formally, let $\dPlayerPoly$
  denote the set of poly-size quantum circuits, so we assume that
  $\dA,\dB \in \dPlayerPoly$. The ideal functionality $\F$ models the
  intended behavior of the protocol in the \emph{ideal world}, where
  the players interact using $\F$. Honest and dishonest players in the
  ideal world (a.k.a.~simulators) are denoted by $\hA, \hB$ and $\dhA,
  \dhB$, respectively. An honest player simply forwards messages to
  and from $\F$, dishonest players are allowed to change their
  messages. Again $\dhA, \dhB \in \dPlayerPoly$. Now, the input-output
  behavior of $\F$ defines the required input-output behavior of
  $\Pi$. Intuitively, if the executions are indistinguishable,
  security of the protocol in real life follows. In other words, a
  dishonest real-world player that attacks protocol $\Pi$ cannot
  achieve (significantly) more than an ideal-world adversary that
  attacks the corresponding functionality $\F$.

  The common input state $\rho_{UV} = \sum_{u,v} P_{UV}(u,v) \proj{u}
  \otimes \proj{v}$ for some probability distribution $P_{UV}$ is
  classical, and we understand $U,V$ as random input variables (for
  Alice and Bob, respectively). The same holds for the classical
  output state $\rho_{X Y}$ with output $X,Y$ for Alice respectively
  Bob.  The input-output behavior of the protocol is uniquely
  determined by $P_{X Y|UV}$, and we write $\Pi(U,V) = (X,Y)$. Then, a
  general classical ideal functionality $\F$ is given by a conditional
  probability distribution $P_{\F(U,V)|UV}$ with $\F(U,V)$ denoting
  the ideal-world execution, where the players forward their inputs
  $U,V$ to $\F$ and output whatever they obtain from $\F$.
  \begin{definition}[Correctness]
    \label{def:correctness}
    A protocol $\Pi(U,V) = (X,Y)$ correctly implements an ideal
    classical functionality $\F$, if for every distribution of the
    input values $U$ and $V$, the resulting common output $(X,Y)$
    satisfies $(U,V,X,Y) \approxs (U,V, \F(U,V))$.
  \end{definition}

  We now define computational security against dishonest Alice, the
  definitions for dishonest Bob are analogue. Let $Z$ and $U'$ denote
  dishonest Alice's classical and quantum information. We consider a
  poly-size quantum circuit, called \emph{input sampler}, which takes
  as input the security parameter and produces the input state
  $\rho_{U' Z V}$. We require from the input sampler that any
  $\rho_{U' Z V}$ is restricted to be of form $ \rho_{\MC{U'}{Z}{V}} =
  \sum_{z,v} P_{ZV}(z,v) \proj{z} \otimes \proj{v} \otimes
  \rho_{U'}^z$ (see~\cite{DFSS07} for notational details), where it
  holds that\footnote{$\rho_E^x$ denotes a state in register $E$,
    depending on value $x \in \X$ of random variable $X$ over $\X$
    with distribution $P_X$. Then, from the view of an observer, who
    holds register $E$ but does not know $X$, the system is in state
    $\rho_E = \sum_{x \in \X} P_X(x) \rho_E^x$, where $\rho_E$ depends
    on $X$ in the sense that $E$ is in state $\rho_E^x$ exactly if $X
    = x$.}  $\rho_{U'}^z = \rho_{U'}^{z,v}$. This expresses
  \emph{conditional independence}, namely that Bob's classical $V$ is
  independent of Alice's quantum part $U'$ when given $Z$. In other
  words, Alice's quantum part~$U'$ is correlated with Bob's part only
  via her classical~$Z$.
  \begin{definition}
    [Computational security against dishonest Alice]
    \label{def:comp.security}
    A protocol $\ \Pi$ implements an ideal classical functionality
    $\F$ computationally securely against dishonest Alice, if for any
    real-world adversary $\dA \in \dPlayerPoly$, there exists an
    ideal-world adversary $\dhA \in \dPlayerPoly$ such that, for any
    efficient input sampler with $\rho_{U' Z V} = \rho_{\MC{U'}{Z}{V}}$,
    it holds that the outputs 
    are
    quantum-computationally indistinguishable, i.e.,
    $
    out_{\dA,\B}^\Pi \approxq out_{\dhA,\hB}^\F \, .
    $
  \end{definition}
  We state these output states explicitly as
  $out_{\dA,\B}^\Pi = \rho_{U X' Z Y}$ and $out_{\dhA,\hB}^\F =
  \rho_{\MC{U X'}{Z}{Y}}$, which shows that Alice's possibilities in the
  ideal world are limited: She can produce some classical input $U$
  for $\F$ from her quantum input state $U'$, and then she can obtain a
  quantum state $X'$ by locally processing $U$ and possibly $\F$'s
  classical reply $X$.


\section{Security Notions for Coin-Flipping}
\label{sec:notions.coin.flip}

\begin{figure}
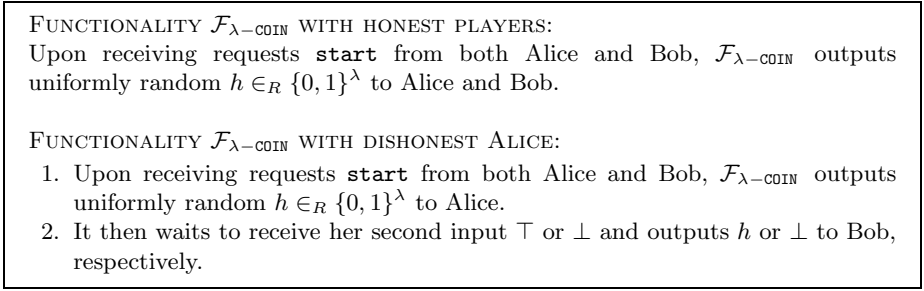

\small
  \begin{framed}\vspace{-1ex}
    \noindent{\sc Functionality} $\cxF{\lambda}$ {\sc
    with honest players: } \\Upon receiving requests $\start$ from both
    Alice and Bob, $\cxF{\lambda}\,$ outputs uniformly random $h \in_R
    \zo^\lambda$ to Alice and Bob.\\\\
    \noindent{\sc Functionality} $\cxF{\lambda}$ {\sc with
    dishonest Alice:}\\[-4.5ex]
    \begin{enumerate}
    \item
      Upon receiving requests $\start$ from both Alice and Bob,
      $\cxF{\lambda}\,$ outputs uniformly random $h \in_R \zo^\lambda$
      to Alice.
    \item
      It then waits to receive her second input $\top$ or $\bot$ and
      outputs $h$ or $\bot$ to Bob, respectively.
    \end{enumerate}
    \vspace{-3ex}
  \end{framed}
  \vspace{-4ex}
  \small
  \caption{The Ideal Functionality for $\lambda$-bit Coin-Flipping.}
  \label{fig:clambda}
\end{figure}

We denote a generic protocol with a $\lambda$-bit coin-string as
output by $\cxPi{\A}{\B}{\lambda}$, corresponding to an ideal
functionality $\cxF{\lambda}$. The outcome of such a protocol is $c
\in \zo^\lambda \cup \set{\bot}$, i.e., either an $\lambda$-bit-string
or an error message. We use several security parameters, indicating
the length of coin-strings for different purposes; the length of a
coin-flip yielding a key or a challenge are denoted by $\kappa$ or
$\sigma$, respectively. The ideal functionality for coin-flipping is
defined symmetric such that always the respective dishonest party has
an option to abort. We state the ideal functionalities in the case of
both players being honest and in the case of dishonest Alice and
honest Bob (Fig.\ \ref{fig:clambda}). Note that the latter then also
applies to honest Alice and dishonest Bob by simply switching sides
and names.

Recall that the \emph{joint output representation} of a protocol
execution is denoted by $out_{\A,\B}^\Pi \,$ (with $\Pi =
\cxPi{\A}{\B}{\lambda}$) and given here for the case of honest
players. The same notation with $\F = \cxF{\lambda}$ and $\hA, \hB$
applies in the ideal world as $out_{\hA,\hB}^\F$, where the players
invoke the ideal functionality $\cxF{\lambda}$ and output whatever
they obtain from it. We need an additional notation here, describing
the \emph{outcome} of a protocol run between e.g.\ honest $\A$ and
$\B$, namely $c \la \cxPi{\A}{\B}{\lambda}$.
 
We will define three flavors of security for coin-flip protocols,
namely \defterm{uncontrollable (uncont)}, \defterm{random} and
\defterm{enforceable (force)}. The two sides can have different
flavors. Then, if a protocol $\cxPi{\A}{\B}{\lambda}$ is, for
instance, enforceable against Alice and random against Bob, we write
$\pi^{(\force,\random)}$, and similarly for the eight other
combinations of security. Note that for simplicity of notation, we
will then omit the indexed name as well as the length of the coin, as
they are clear from the context. Again, we define all three flavors
for Alice's side only, as the definitions for Bob are analogue. Recall
that $U'$ and $Z$ resp.\ $V$ denote dishonest Alice's quantum and
classical input resp.\ honest Bob's classical input. As before, we
assume a poly-size input sampler, which takes as input the security
parameter, and produces a valid input state $\rho_{U' Z V} =
\rho_{\MC{U'}{Z}{V}}$. Note that an honest player's input is empty but
models the invocation $\start$. We stress that we require for all
three security flavors and for all $c \in \zo^{\lambda}$ that 
$$\prob{c \la \cxPi{\A}{\B}{\lambda}} = 2^{-\lambda} \, ,$$ which
implies that when both parties are honest, then the coin is
unbiased. Below we only define the extra properties required for each
of the three flavors.

We call a coin-flip \defterm{uncontrollable} against Alice, if she
cannot force the coin to hit some negligible subset, except with
negligible probability.
\begin{definition}[Uncontrollability against dishonest Alice]
  \label{def:uncont}
  We say that protocol $\cxPi{\A}{\B}{\lambda}$ implements an
  \defterm{uncontrollable} coin-flip against dishonest Alice, if it
  holds for any poly-sized adversary $\dA \in \dPlayerPoly$ with inputs
  as specified above and all negligible subsets $Q \subset
  \zo^\lambda$ that
    $$ \prob{c \la \cxPi{\dA}{\B}{\lambda}\, : \, c \in Q} \in
    \negl{\kappa} \, .
    $$
  \end{definition}
Note that we denote by $Q \subset \zo^\lambda$ a family of subsets
$\set{Q(\kappa) \subset \zo^{\lambda(\kappa)}}_{\kappa \in \NN}$ for
security parameter $\kappa$. Then we call $Q$ negligible, if $\vert
Q(\kappa) \vert 2^{-\lambda(\kappa)}$ is negligible in $\kappa$. In
other words, we call a subset negligible, if it contains a negligible
fraction of the elements in the set in which it lives.

We call a coin-flip \defterm{random} against Alice, if she cannot
enforce a non-uniformly random output string in $\zo^\lambda$, except
by making the protocol fail on some chosen runs. That means she can at
most lower the probability of certain output strings compared to the
uniform case.
\begin{definition}[Randomness against dishonest Alice]
  \label{def:random}
  We say that $ $ protocol $\ \cxPi{\A}{\B}{\lambda}$ implements a
  \defterm{random} coin-flip against dishonest Alice, if it holds for
  any poly-sized adversary $\dA \in \dPlayerPoly$ with inputs as
  specified above that there exists an event E such that $\prob{E} \in
  \negl{\kappa}$ and for all $x \in \zo^\lambda$ it holds that
    $$
    \prob{c \la
    \cxPi{\dA}{\B}{\lambda} \, : \, c = x \, \vert \, \bar{E}} \leq
    2^{-\lambda} \, .
    $$
\end{definition}
It is obvious that if a coin-flip is random against Alice, then it is
also an uncontrollable coin-flip against her. We will later discuss a
generic transformation going in the other direction from
uncontrollable to random coin-flipping.

We call a coin-flip \defterm{enforceable} against Alice, if it is
possible, given a uniformly random $c$, to simulate a run of the
protocol hitting exactly the outcome $c$, though we still allow that
the corrupted party forces abort on some outcomes.\footnote{Note that
  an enforceable coin-flip is not necessarily a random coin-flip, as
  it is allowed that the outcome of an enforceable coin-flip is only
  quantum-computationally indistinguishable from uniformly random,
  whereas a random coin-flip is required to produce truly random
  outcomes on the non-aborting runs.}
\begin{definition}[Enforceability against dishonest Alice]
  \label{def:force}
  We call protocol $\cxPi{\A}{\B}{\lambda}$ \defterm{enforceable}
  against dishonest Alice, if it implements the ideal functionality
  $\cxF{\lambda}$ against her.
\end{definition}
That means that for any poly-sized adversary $\dA \in \dPlayerPoly$,
there exists an ideal-world adversary $\dhA \in \dPlayerPoly$ that
simulates the protocol with $\dA$ as follows. $\dhA$ requests output
$h \in \zo^\lambda$ from $\cxF{\lambda}$. Then it simulates a run of
the coin-flip protocol with $\dA$ and tries to enforce output $h$. If
$\dhA$ succeeds, it inputs $\top$ as $\dA$'s second input to
$\cxF{\lambda}$. In that case, $\cxF{\lambda}$ outputs $h$. Otherwise,
$\dhA$ inputs $\bot$ to $\cxF{\lambda}$ as second input and
$\cxF{\lambda}$ outputs $\bot$. In addition, the simulation is such
that the ideal output is quantum-computationally indistinguishable
from the output of an actual run of the protocol, i.e.,
$out_{\dA,\B}^\Pi \approxq out_{\dhA,\hB}^\F$, where $\Pi =
\cxPi{\dA}{\B}{\lambda}$ and $\F = \cxF{\lambda}$. Enforceability
against dishonest Bob is analogously
defined. Corollary~\ref{cor:double.simulatable} follows.
\begin{corollary}
  \label{cor:double.simulatable}
  If $\cxPi{\A}{\B}{\lambda} \in \pi^{(\force,\force)}$, i.e., it is
  enforceable against both dishonest Alice and dishonest Bob, then
  $\cxPi{\A}{\B}{\lambda}$ is a secure implementation of
  $\cxF{\lambda}$, according to Definition~\ref{def:comp.security}.
\end{corollary}


\section{Mixed Commitments}
\label{sec:mixed.commit}

We use mixed commitment schemes throughout our constructions---they
will indeed be our only computational assumption. Mixed commitment are
unconditionally hiding for some public keys and unconditionally
binding for others. In the following, we introduce mixed commitments,
denoted by $\commit_{pk}$, more formally. We also describe a
construction of an interactive commitment protocol $\Commit_{pk}$ with
mixed-commitment-scheme-like properties. The reason for presenting the
protocol here is to simplify the description of the later protocol in
which it is used as a subprotocol.


\subsection{Mixed Commitment Schemes}
\label{sec:mixed}

  Mixed commitment schemes consists of four poly-time algorithms $
  \GH$, $\GB$, $\commit$, and $\xtrx$. The \defterm{unconditionally
    hiding key generator} $\GH$ outputs public keys $pk \in
  \zo^\kappa$.\footnote{For notational simplicity, the length of
    public keys is assumed to equal security parameter $\kappa$.}  The
  \defterm{unconditionally binding key generator} $\GB$ outputs key
  pairs $(pk,sk)$, where $pk \in \zo^\kappa$ and where $sk$ is the
  secret key. The commitment algorithm takes as input a message $m$, a
  randomizer $r$ and a public key $pk$ and outputs a commitment $C =
  \commitk{m}{r}{pk}$. The extraction algorithm $\xtrx$ takes as input
  a commitment $C$ and a secret key $sk$ and outputs a message $m'$,
  meant to be the message committed by $C$. We require the following
  properties:\\\\
  \noindent{\bf Unconditionally hiding: }For keys $pk$ generated by
  $\GH$ it holds that $\commit_{pk}$ is statistically hiding,
  i.e.\ $(pk,\commitk{m_1}{r_1}{pk}) \approxs
  (pk,\commitk{m_2}{r_2}{pk})$ for all $m_1, m_2$ when $r_1$ and $r_2$
  are uniformly random and independent.\\
  \noindent {\bf Extractability: } It holds for all pairs $(pk,sk)$
  generated by $\GB$ and for all values $m,r$ that
  $\xtr{\commitk{m}{r}{pk}}{sk} = m$. \\
  \noindent{\bf Key indistinguishability: } A random public key $pk_1$
  generated by $\GB$ and a random public key $pk_2$ generated by $\GH$
  are indistinguishable by poly-sized quantum circuits, i.e., $pk_1
  \approxq pk_2$.\\

  We additionally require that random public keys generated by $\GH$
  are statistically close to uniform in $\zo^\kappa$, i.e., almost all
  keys are unconditionally hiding.\footnote{The definition is a
    weakening of the original notion of mixed commitments
    from~\cite{DN02}, in that we do not require that unconditionally
    hiding keys are equipped with an equivocation trapdoor. It is also
    a strengthening in that we require quantum indistinguishability of
    the two key flavors.}
 
  As a candidate for instantiating our definition we can, for
  instance, take the lattice-based public-key encryption scheme of
  Regev~\cite{Regev05} in its multi-bit variant as given in the full
  version of~\cite{PVW08}. Regev's cryptosystem is based on the
  hardness of the learning with error problem, which can be reduced
  from worst-case (quantum) hardness of the shortest vector problem
  (in its decision version). Thus, breaking the scheme implies an
  efficient algorithm for approximating the lattice problem in the
  worst-case, which is assumed to be hard even with quantum computing
  power. A regular public key for Regev's scheme is proven to be
  quantum-computationally indistinguishable from the case where a
  public key is chosen from the uniform distribution. In this case,
  the ciphertext carries essentially no information about the
  message~\cite[Lemma 5.4]{Regev05}. This proof of semantic security
  for Regev's cryptosystem is in fact the property we require for our
  commitment.


\subsection{The protocol $\Commit_{pk}$}
\label{sec:Mixed}

  In one of our security amplifications of coin-flip protocols we will
  need a mixed commitment scheme which also provides
  \emph{equivocability}, i.e., a simulator can open unconditionally
  hiding commitments to different values. We add equivocability using
  an interactive protocol $\Commit_{pk}$. Instead of equipping
  unconditionally hiding keys with equivocation trapdoors, we will do
  it by letting the equivocation trapdoor be the ability of the
  simulator to force the outcome of a coin-flip protocol in the
  simulation. The reason for this change, as compared to~\cite{DN02},
  is that the notion of a mixed commitment scheme in~\cite{DN02} was
  developed for the CRS-model, where the simulator is free to pick the
  CRS and hence could pick it to be a unconditionally hiding public
  key with known equivocation trapdoor. Here we are interested in the
  bare (CRS devoid) model and hence have to add equivocation in a
  different manner. This is one of the essential steps in
  bootstrapping fully simulatable strong coin-flipping from weak
  coin-flipping.

  The protocol $\Commit_{pk}$ uses a secret sharing scheme $\sss$,
  described now. Let $\sigma$ be a secondary security parameter. Given
  message $m = (m_1,\ldots,m_\sigma) \in \FF^\sigma$ and randomizer $s
  = (s_1, \ldots, s_\sigma) \in \FF^\sigma$, let $f_{m,s}({\tt X})$
  denote the unique polynomial of degree $2\sigma-1$, for which
  $f_{m,s}(-i+1) = m_i$ for $i = 1, \ldots, \sigma$ and $f_{m,s}(i) =
  s_i$ for $i = 1, \ldots, \sigma$. Furthermore, we ``fill up''
  positions $i = \sigma+1, \ldots, \Sigma$, where $\Sigma = 4 \sigma$,
  by letting $s_i = f_{m,s}(i)$. The shares are now $s =
  (s_1,\ldots,s_\Sigma)$.

  We stress two simple facts about $\sss$. First, for any message $m
  \in \FF^\sigma$ and any subset $S \subset \{ 1, \ldots, \Sigma \}$
  of size $\vert S \vert = \sigma$, the shares $s|_S$ are uniformly
  random in $\FF^\sigma$, when $S$ is chosen uniformly at random in
  $\FF^\sigma$ and independent of $m$. This aspect is trivial for $S =
  \{ 1, \ldots, \sigma \}$, as we defined it that way, and it extends
  to the other subsets using Lagrange interpolation. And second, if
  $m^1, m^2 \in \FF^\sigma$ are two distinct messages, then
  $\sss(m^1;s^1)$ and $\sss(m^2;s^2)$ have Hamming distance at least
  $\Sigma - 2 \sigma$. Again, this follows by Lagrange interpolation,
  since the polynomial $f_{m^1,s^1}({\tt X})$ has degree at most
  $2\sigma-1$, and hence, can be computed from any $2 \sigma$ shares
  $s_i$ using Lagrange interpolation. The same holds for
  $f_{m^2,s^2}({\tt X})$. Thus, if $2\sigma$ shares are the same, then
  $f_{m^1,s^1}({\tt X})$ and $f_{m^2,s^2}({\tt X})$ are the same,
  which implies that the messages $m^1 =
  f_{m^1,s^1}(-\sigma+1),\ldots, f_{m^1,s^1}(0)$ and $m^2 =
  f_{m^2,s^2}(-\sigma+1),\ldots, f_{m^2,s^2}(0)$ are the same.

\begin{figure}
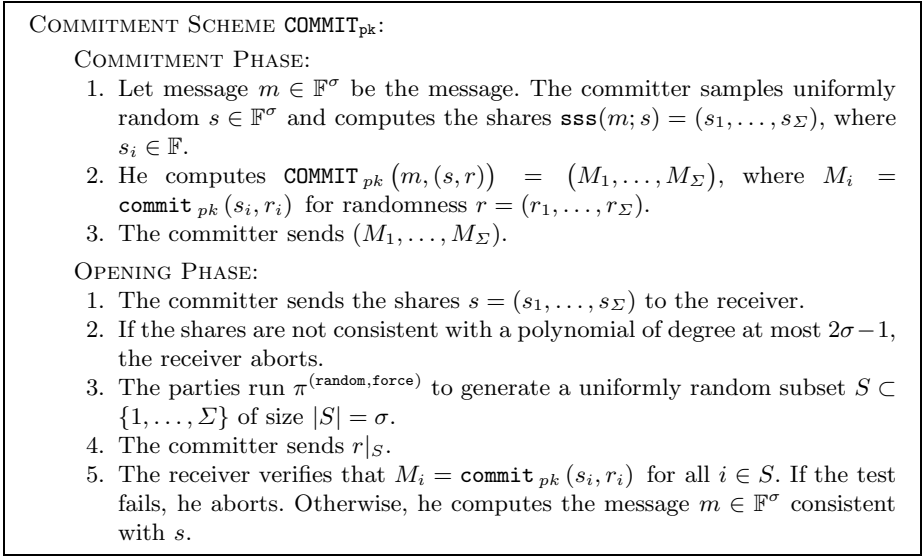

   \small
    \begin{framed}\vspace{-1ex}
      \noindent{\sc Commitment Scheme $\Commit_{\pk}$:}
      \begin{itemize}\vspace{-1.5ex}
      \item[]{\sc Commitment Phase:}
	\begin{enumerate}
	\item
	  Let message $m \in \FF^\sigma$ be the message.
	  The committer samples uniformly random $s \in
	  \FF^\sigma$ and computes the shares $\sss(m;s) = (s_1,
	  \ldots, s_\Sigma)$, where $s_i \in \FF$.
	\item
	  He computes $\Commitk{m}{(s,r)}{pk} = \big(
	  M_1,\ldots,M_\Sigma \big)$, where $M_i =
	  \commitk{s_i}{r_i}{pk}$ for randomness $r =
	  (r_1,\ldots,r_\Sigma)$.
	\item
	  The committer sends $(M_1,\ldots,M_\Sigma)$.\\[-2ex]
	\end{enumerate}
      \item[]{\sc Opening Phase:}
	\begin{enumerate}
	\item
	  The committer sends the shares $s = (s_1, \ldots, s_\Sigma)$
	  to the receiver.
	\item
	  If the shares are not consistent with a polynomial of degree
	  at most $2\sigma-1$, the receiver aborts.
	\item
	  The parties run
	  $\pi^{(\random,\force)}$ to generate a uniformly random
	  subset $S \subset \set{1,\ldots,\Sigma}$ of size $\vert S
	  \vert = \sigma$.
	\item
	  The committer sends $r|_S$.
	\item
	  The receiver verifies that $M_i = \commitk{s_i}{r_i}{pk}$
	  for all $i \in S$. If the test fails, he aborts. Otherwise,
	  he computes the message $m \in \FF^\sigma$ consistent with
	  $s$.
	\end{enumerate}
      \end{itemize}
      \vspace{-3ex}
    \end{framed}
    \vspace{-4ex}
    \small
    \caption{The Commitment Scheme $\Commit_{pk}$.}
    \label{fig:sss.commit}
  \end{figure}

  In addition to $\sss$, the protocol $\Commit_{pk}$ uses a mixed
  commitment scheme $\commit_{pk}$. The key generators for
  $\Commit_{pk}$ are the same as for $\commit_{pk}$. Finally,
  $\Commit_{pk}$ uses a coin-flip protocol $\pi^{(\random,\force)}$
  which is random for the committer and which is enforceable against
  the receiver of the commitment. The details of $\Commit_{pk}$ are
  given in Fig.\ \ref{fig:sss.commit}.

  We first show that when $(pk,sk)$ is generated using $\GB$, then
  $\Commit_{pk}$ is extractable. Given any commitment $M = \big(
  M_1,\ldots,M_\Sigma \big)$, we extract
  $
  \xtr{M}{sk} = \big( \xtr{M_1}{sk},\ldots,\xtr{M_\Sigma}{sk} \big) =
  (s_1,\ldots,s_\Sigma) = s \, .
  $
  Assume $s' = (s'_1,\ldots,s'_\Sigma)$ is the consistent sharing
  closest to $s$. That means that $s'$ is the vector which is
  consistent with a polynomial $f_{m',s'}({\tt X})$ of degree at most
  $2\sigma-1$ and which at the same time differs from $s$ in the
  fewest positions. Note that we can find $s'$ in poly-time when using
  a Reed Solomon code, which has efficient minimal distance
  decoding. We then interpolate the polynomial $f_{m',s'}({\tt X})$,
  let $m' = f_{m',s'}(-\sigma+1),\ldots,f_{m',s'}(0)$, and let
  $\xtr{M}{sk} = m'$. Any other sharing $s''=
  (s_1'',\ldots,s_\Sigma'')$ must have Hamming distance at least $2
  \sigma$ to $s'$. Now, since $s$ is closer to $s'$ than to any other
  consistent sharing, it must, in particular, be closer to $s'$ then
  to $s''$. This implies that $s$ is at distance at least $\sigma$ to
  $s''$.

  We will use this observation for proving soundness of the opening
  phase. To determine the soundness error, assume that $\Commit_{pk}$
  does not open to the shares $s'$ consistent with $s$. As observed,
  this implies that $\big( \xtr{M_1}{sk},\ldots,\xtr{M_\Sigma}{sk}
  \big)$ has Hamming distance at least $\sigma$ to $s'$. However, when
  $\commit_{pk}$ is unconditionally binding, all $M_i$ can only be
  opened to $\xtr{M_i}{sk}$. From the above two facts, we have that
  there are at least $\sigma$ values $i \in \set{1,\ldots,\Sigma}$
  such that the receiver cannot open $M_i$ to $s_i$ for $i \in
  S$. Since $\Sigma = 4 \sigma$, these $\sigma$ bad indices (bad for a
  dishonest sender) account for a fraction of $\frac14$ of all points
  in $\set{1,\ldots,\Sigma}$. Thus, the probability that none of the
  $\sigma$ points in $S$ is a bad index is at most $(\frac34)^\sigma$,
  which is negligible. Setting $\sigma = \log_{\frac43}2$ gives a
  negligible error of $(\frac12)^\kappa$, where $\kappa$ is the
  security parameter.

  We then analyze the equivocability of $\Commit_{pk}$. We will use
  the ability of the simulator for the committer to force the
  challenge $S$ as the simulator's trapdoor. It will simply pick $S$
  uniformly at random before the simulation and prepare for this
  particular challenge. The details are given in
  Fig.\ \ref{fig:simulation.sss}. We omit an analysis here but refer
  to Section~\ref{sec:amplification.force.force}, where the
  construction will be further discussed.

  \begin{figure}
    \small
    \begin{framed}\vspace{-1ex}
      \noindent{\sc Simulating $\Commit_{pk}$ with
	Trapdoor $S$:} \\[-4.5ex]
      \begin{enumerate}
      \item
	$\hat{\Sim}$ gets as input a uniformly random subset $S
	\subset \set{1,\ldots,\Sigma}$ of size $\sigma$ and an initial
	message $m \in \FF^\sigma$.
      \item
	$\hat{\Sim}$ commits honestly to $m \in \FF^\sigma$ by $M =
	\Commitk{m}{(s,r)}{sk}$, as specified in the commitment phase.
      \item
	$\hat{\Sim}$ is given an alternative message $\tilde{m} \in
	\FF^\sigma$, i.e., the aim is opening $M$ to $\tilde{m}$.
      \item
	$\hat{\Sim}$ lets $s|_S$ be the $\sigma$ messages committed to
	by $M|_S$. Then it interpolates the unique polynomial
	$f_{\tilde{m},s}$ of degree at most $2\sigma-1$ for which
	$f_{\tilde{m},s}(i) = s_i$ for $i \in S$ and for which
	$f_{\tilde{m},s}(-i+1) = \tilde{m}_i$ for $i =
	1,\ldots,\sigma$. Note that this is possible, as we have
	exactly $2\sigma$ points which restrict our choice of
	$f_{\tilde{m},s}$. $\hat{\Sim}$ sends $s = \big(
	f_{\tilde{m},s}(1),\ldots,f_{\tilde{m},s}(\Sigma) \big)$ to
	the receiver.
      \item
	The parties run $\pi^{(\random,\force)}$ and $\hat{\Sim}$
	forces the outcome $S$.
      \item
	For all $i \in S$, the sender opens $M_i$ to
	$f_{\tilde{m},s}(i)$. This is possible, since
	$f_{\tilde{m},s}(i) = s_i$ is exactly the message committed to
	by $M_i$ when $i \in S$.
      \end{enumerate}
      \vspace{-3ex}
    \end{framed}
    \vspace{-4ex}
    \small
    \caption{The Ideal-World Simulation of $\Commit_{pk}$.}
    \label{fig:simulation.sss}
  \end{figure}


\section{Amplification Theorems for Strong Coin-Flipping}
\label{sec:amplification.coin.flip}

We now propose and prove theorems, which allow us to amplify the
security strength of coins. Ultimately, we aim at constructing a
strong coin-flip protocol $\pi^{(\force,\force)}$ with outcomes of any
polynomial length $\ell$ in $\lambda$ from a weaker coin-flip protocol
$\pi^{(\force,\uncont)}$ of $\kappa$-bit-strings, where $\kappa$ is
the key length of the mixed commitment scheme. We do this in two
steps. We first show how to implement $\pi^{(\force,\random)}$ for
$\ell$-bit-strings (for any polynomial $\ell$) given
$\pi^{(\force,\uncont)}$ for $\kappa$-bit-strings, and we then show
how to implement $\pi^{(\force,\force)}$ for poly-long bit-strings
given $\pi^{(\force,\random)}$ for poly-long bit-strings.

The ability to amplify $\pi^{(\force,\uncont)}$ for
$\kappa$-bit-strings to $\pi^{(\force,\force)}$ for poly-bit-string is
of course only interesting, if there exists such a candidate. We do
not know of any protocol with flavor $(\force,\uncont)$ but not
$(\force,\random)$. However, we consider it as a contribution in
itself to find the weakest security notion for coin-flipping that
allows to amplify to the final strong $(\force,\force)$ notion using a
constant-round reduction.

A candidate for $\pi^{(\force,\random)}$ with one-bit outcomes is the
protocol in~\cite{DL09}, which is---in terms of this
context---enforceable against one side in poly-time and random on the
other side, with empty event $E$ according to
Definition~\ref{def:random}, and the randomness guarantee even
withstanding an unbounded adversary.\footnote{The protocol was
described and proven as $\pi^{(\random,\force)}$, but due to the
symmetric coin-flip definitions here, we can easily switch sides
between $\A$ and $\B$.} The protocol was shown to be sequentially
composable~\cite{DL09,Lunemann10}. Repeating the protocol $\kappa$
times in sequence gives a protocol $\pi^{(\force,\random)}$ for
$\kappa$-bit-strings. Note that this, in particular, gives a protocol
$\pi^{(\force,\uncont)}$ for $\kappa$-bit-strings.


\subsection{From $(\force,\uncont)$ to $(\force,\random)$}
\label{sec:amplification.uncont.random}

  Assume that we are given a protocol $\pi^{(\force,\uncont)}$, that
  only guarantees that Bob cannot force the coin to hit a negligible
  subset (except with negligible probability). We now amplify the
  security on Bob's side from $\defterm{uncontrollable}$ to
  $\defterm{random}$ and therewith obtain a protocol
  $\pi^{(\force,\random)}$, in which Bob cannot enforce a
  non-uniformly random output string, except by letting the protocol
  fail on some occasions. The stronger protocol
  $\pi^{(\force,\random)}$ is given in Fig.\ \ref{fig:force.random},
  where $\commit_{pk}$ is the basic mixed commitment scheme as
  described in Section~\ref{sec:mixed}. Correctness of
  $\pi^{(\force,\random)}$ is obvious by inspection of the
  protocol.

  \begin{figure}
    \small
    \begin{framed}\vspace{-1ex}
      \noindent{\sc Protocol}
      $\pi^{(\force,\random)}$: \\[-4.5ex]
      \begin{enumerate}
      \item
	$\A$ and $\B$ run $\pi^{(\force,\uncont)}$ to produce a public
	key $pk \in \zo^\kappa$.
      \item
	$\A$ samples $a \in_R \zo^\ell$, commits to it with $A =
	\commitk{a}{r}{pk}$ and randomizer $r \in_R \zo^\ell$, and
	sends $A$ to $\B$.
      \item
	$\B$ samples $b \in_R \zo^\ell$ and sends $b$ to $\A$.
      \item
      $\A$ opens $A$ towards $\B$.
      \item
	The outcome is $c = a \oplus b$.
      \end{enumerate}
      \vspace{-3ex}
    \end{framed}
    \vspace{-4ex}
    \small
    \caption{Amplification from $(\force,\uncont)$ to $(\force,\random)$.}
    \label{fig:force.random}
  \end{figure}

  \begin{theorem}
    \label{thm:force.random}
    If $\pi^{(\force,\uncont)}$ is enforceable against Alice and
    uncontrollable \\ against Bob, then protocol $\pi^{(\force,\random)}$
    is enforceable against Alice and random for Bob.
  \end{theorem}
  We sketch the basic ideas behind the proof, which can be found in
  greater detail in Appendix~\ref{app:proof.force.random}. Enforceability against $\A$ follows by forcing
  $pk$ to be a $pk$ generated as $(pk,sk) \leftarrow \GB$. The
  simulator then uses $sk$ to extract $a$ from $A$ and then sends the
  $b$ which makes $a \oplus b$ hit the desired outcome. Randomness
  against $\B$ follows from the fact that only a negligible fraction
  of the keys $pk \in \zo^\kappa$ are not unconditionally hiding keys
  and the outcome of $\pi^{(\force,\uncont)}$ is uncontrollable for
  $\B$.


\subsection{From $(\force,\random)$ to $(\force,\force)$}
\label{sec:amplification.force.force}

\begin{figure}
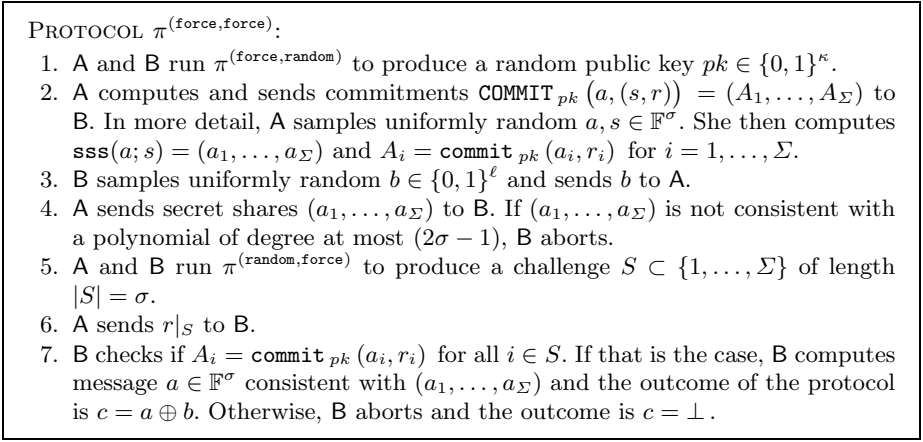

    \small
    \begin{framed}\vspace{-1ex}
      \noindent{\sc Protocol} $\pi^{(\force,\force)}$:
      \\[-4.5ex]
      \begin{enumerate}
      \item
	$\A$ and $\B$ run $\pi^{(\force,\random)}$ to produce a random
	public key $pk \in \zo^\kappa$.
      \item
	$\A$ computes and sends commitments $\Commitk{a}{(s,r)}{pk} =
	(A_1,\ldots,A_\Sigma) $ to $\B$. In more detail, $\A$ samples
	uniformly random $a, s \in \FF^\sigma$. She then computes
	$\sss(a;s) = (a_1,\ldots,a_\Sigma)$ and $A_i =
	\commitk{a_i}{r_i}{pk}$ for $i = 1, \ldots, \Sigma$.
      \item
	$\B$ samples uniformly random $b \in \zo^\ell$ and sends $b$ to
	$\A$.
      \item
	$\A$ sends secret shares $(a_1,\ldots,a_\Sigma)$ to $\B$. If
	$(a_1, \ldots, a_\Sigma)$ is not consistent with a polynomial of
	degree at most $(2\sigma-1)$, $\B$ aborts.
      \item
	$\A$ and $\B$ run $\pi^{(\random,\force)}$ to produce a
	challenge $S \subset \set{1,\ldots,\Sigma}$ of length $\vert S
	\vert = \sigma$.
      \item
	$\A$ sends $r|_S$ to $\B$.
      \item
	$\B$ checks if $A_i = \commitk{a_i}{r_i}{pk}$ for all $i \in
	S$. If that is the case, $\B$ computes message $a \in
	\FF^\sigma$ consistent with $(a_1, \ldots, a_\Sigma)$ and the
	outcome of the protocol is $c = a \oplus b$. Otherwise, $\B$
	aborts and the outcome is $c = \bot\,$.
      \end{enumerate}
      \vspace{-3ex}
    \end{framed}
    \vspace{-4ex}
    \small
    \caption{Amplification from $(\force,\random)$ to
      $(\force,\force)$.}
    \label{fig:force.force}
  \end{figure}

  We now show how to obtain a coin-flip protocol, which is enforceable
  against both parties. Then, we can also claim by
  Corollary~\ref{cor:double.simulatable} that this protocol is a
  strong coin-flip protocol, poly-time simulatable on both sides for
  the natural ideal functionality $\clF$. The protocol
  $\pi^{(\force,\force)}$ is described in Fig.\ \ref{fig:force.force}
  and uses the extended commitment construction $\Commit_{pk}$ from
  Section~\ref{sec:Mixed}. The protocol makes two calls to a
  subprotocol with random flavor on one side and enforceability on the
  other side, but where the sides are interchanged,
  i.e.~$\pi^{(\force,\random)}$ and $\pi^{(\random,\force)}$, so we
  simply switch the players' roles. Again, correctness of the protocol
  can be trivially checked.

  \begin{theorem}
    \label{thm:force.force}
    If $\pi^{(\force,\random)}$ is enforceable against Alice and
    random against Bob, then protocol $\pi^{(\force,\force)}$ is
    enforceable against both Alice and Bob.
  \end{theorem}
  We sketch the main ideas behind the proof, which can be found in
  greater detail in Appendix~\ref{app:proof.force.force}. Enforceability against $\A$ follows by forcing
  $pk$ to be a key $pk$ generated as $(pk,sk) \leftarrow \GB$. The
  simulator then uses $sk$ to extract $a$ from $(A_1, \ldots,
  A_\Sigma)$. Then it sends the $b$ that makes $a \oplus b$ hit the
  desired outcome. Enforceability against $\B$ follows by letting the
  simulator sample a uniformly random $S$ and running
  $\Commitk{a}{(s,r)}{pk} = (A_1,\ldots,A_\Sigma)$ in the equivocal
  model with trapdoor $S$. Then the simulator waits for $b$ and forces
  the outcome of $\pi^{(\random,\force)}$ to be $S$, which allows it
  to open $(A_1, \ldots, A_\Sigma)$ to the $a$ that makes $a \oplus b$
  hit the desired outcome.


\section{Application: Zero-Knowledge Proof of Knowledge}
\label{sec:coin.zkpk}

The purpose of a zero-knowledge proof of knowledge~\cite{GMR85,BG93}
is to verify in classical poly-time in the length of the instance,
whether the prover's private input $w$ is a valid witness for the
common instance $x$ in relation $\Rel$, i.e.~$(x,w) \in \Rel$. Here,
we propose a quantum-secure construction of a zero-knowledge proof of
knowledge based on witness encoding, which we define in the context of
a simulation in the quantum world. The protocol is constant-round if
the coin-flip protocol is constant-round.


\subsection{Simulatable Witness Encodings of \NP}
\label{sec:zkpk.witness.encoding}

  We first specify a simulatable encoding scheme for binary relation
  $\Rel \subset \zo^* \times \zo^*$, which consists of five classical
  poly-time algorithms $(E,D,S,J,\hat{E})$. Then, we define
  completeness, extractability and simulatability for such a scheme in
  terms of the requirements of our zero-knowledge proof of knowledge.

  Let $E: \Rel \times \zo^m \ra \zo^n$ denote an \defterm{encoder},
  such that for each $(x,w) \in \Rel$, the $n$-bit output $e \la
  E(x,w,r')$ is a random encoding of $w$, with randomness $r' \in
  \zo^m$ and polynomials $m(\vert x \vert)$ and $n(\vert x
  \vert)$. The corresponding \defterm{decoder} $D: \zo^* \times \zo^n
  \ra \zo^*$ takes as input an instance $x \in \zo^*$ and an encoding
  $e \in \zo^n$ and outputs $w \la D(x,e)$ with $w \in \zo^*$. Next,
  let $S$ denote a \defterm{selector} with input $s \in \zo^\sigma$
  (with polynomial $\sigma(\vert x \vert)$) specifying a challenge,
  and output $S(s)$ defining a poly-sized subset of $\set{1,\ldots,n}$
  corresponding to challenge $s$. We will use $S(s)$ to select which
  bits of an encoding $e$ to reveal to the verifier. For simplicity,
  we use $e_s$ to denote the collection of bits $e|_{S(s)}$. We denote
  with $J$ the \defterm{judgment} that checks a potential encoding $e$
  by inspecting only bits $e_s$. In more detail, $J$ takes as input
  instance $x \in \zo^*$, challenge $s \in \zo^\sigma$ and the $\vert
  S(s) \vert$ bits $e_s$, and outputs a judgment $j \la J(x,s,e_s)$
  with $j \in \set{\abort, \success}$. Finally, the
  \defterm{simulator} is called $\hat{E}$. It takes as input instance
  $x \in \zo^*$ and challenge $s \in \zo^\sigma$ and outputs a random
  collection of bits $t|_{S(s)} \la \hat{E}(x,s)$. Again for
  simplicity, we let $t_s = t|_{S(s)}$. Then, if this set has the same
  distribution as bits of an encoding $e$ in positions $S(s)$, the
  bits needed for the judgment to check an encoding $e$ can be
  simulated given just instance $x$ (see
  Definition~\ref{def:simulate}).
  \begin{definition}[Completeness]
    \label{def:zkpk.complete}
    If an encoding $e\la~E(x,w,r)$ is generated correctly, then
    $\success \la J(x,s,e_s)$ for all $s \in \zo^\sigma$.
  \end{definition}
 
  We will call an encoding $e$ \defterm{admissible} for $x$, if there
  \emph{exist} two distinct challenges $s,s' \in \zo^\sigma$ for which
  $\success \la J(x,s,e_s)$ and $\success \la J(x,s',e_{s'})$.
  \begin{definition}[Extractability]
    \label{def:extract}
    If an encoding $e$ is admissible for $x$, then $\big( x,D(x,e)
    \big)\in \Rel$.
  \end{definition}
  We stress that extractability is similarly defined to the
  special soundness property of a classical $\Sigma$-protocol, which
  allows to extract $w$ from two accepting conversations with
  distinct challenges. Such a requirement would generally be
  inapplicable in the quantum setting, as the usual rewinding
  technique is problematic and in particular in the context here, we
  cannot measure two accepting conversations during rewinding in the
  quantum world. Therefore, we define the stronger requirement that if
  there \emph{exist} two distinct answerable challenges for one
  encoding $e$, then $w$ can be extracted given only $e$. This
  condition works nicely in the quantum world, since we can obtain $e$
  without rewinding, as we demonstrate below.
  \begin{definition}[Simulatability]
    \label{def:simulate}
    For all $(x,w) \in \Rel$ and all $s \in_R \zo^\sigma$, the
    distribution of $e \leftarrow E(x,w,r')$ restricted to positions
    $S(s)$ is identical to the distribution of $t_s \leftarrow
    \hat{E}(x,s)$.
  \end{definition}
 
  To construct a simulatable witness encoding one can, for instance,
  start from the commit-and-open protocol for circuit satisfiability
  in~\cite{BCC88}, where the bits of the randomized circuit committed
  to by the sender is easy to see as a simulatable encoding of a
  witness being a consistent evaluation of the circuit to output
  $1$. The challenge in the protocol is one bit $e$ and the prover
  replies by showing either the bits corresponding to some positions
  $S'(0)$ or positions $S'(1)$. The details can be found
  in~\cite{BCC88}. This gives us a simulatable witness encoding for
  any $\NP$-relation $\Rel$ with $\sigma = 1$, using a Karp reduction
  from $\NP$ to circuit simulatability. By repeating it $\sigma$ times
  in parallel we get a simulatable witness encoding for any
  $\sigma$. For $i = 1, \ldots, \sigma$, compute an encoding $e^i$ of
  $w$ and let $e = (e^1, \ldots, e^\sigma)$. Then for $s \in
  \zo^\sigma$, let $S(s)$ specify that the bits $S'(s_i)$ should be
  shown in $e^i$ and check these bits. Note, in particular, that if
  two distinct $s$ and $s'$ passes this judgment, then there exists
  $i$ such that $s_i \ne s_i'$, so $e^i$ passes the judgment for both
  $s_i = 0$ and $s_i = 1$, which by the properties of the protocol for
  circuit satisfiability allows to compute a witness $w$ for $x$ from
  $e^i$. One can find $w$ from $e$ simply by trying to decode each
  $e^j$ for $j = 1, \ldots, \sigma$ and check if $(x,w_j) \in \Rel$.


\subsection{The Protocol}
\label{sec:zkpk.protocol}

  We now construct a quantum-secure zero-knowledge proof of knowledge
  from prover $\A$ to verifier $\B$. We are interested in the
  \NP-language \\${\cal L(R)} = \set{ x \in \zo^* \, \vert \, \exists \,
  w \ \text{s.t.} \ (x,w) \in \Rel }$, where $\A$ has input $x$ and
  $w$, and both $\A$ and $\B$ receive positive or negative judgment of
  the validity of the proof as output. We assume in the following that
  on input $(x,w) \notin \Rel$, honest $\A$ aborts. Unlike
  zero-knowledge proofs, proofs of knowledge can be modeled by an
  ideal functionality, given as $\zkpkF$ in Fig.\
  \ref{fig:zkpkF}. $\zkpkF$ can be thought of as a channel which only
  allows to send messages in the language $\cal L(R)$. It models
  \emph{zero-knowledge}, as it only leaks instance $x$ and judgment
  $j$ but not witness $w$. Furthermore, it models a \emph{proof of
  knowledge}, since Alice has to know and input a valid witness $w$ to
  obtain output $j = \success$.

  \begin{figure}
    \small
    \begin{framed}\vspace{-1ex}
      \noindent{\sc Functionality $\zkpkF$:}\\[-4.5ex]
      \begin{enumerate}
      \item
	On input $(x,w)$ from Alice, $\zkpkF$ sets $j = \success$ if
	$(x,w) \in \Rel$. Otherwise, it sets $j = \abort$.
      \item
	$\zkpkF$ outputs $(x,j)$ to Bob.
      \end{enumerate}
      \vspace{-3ex}
    \end{framed}
    \vspace{-4ex}
    \small
    \caption{The Ideal Functionality for a Zero-Knowledge Proof of
      Knowledge.}\label{fig:zkpkF}
  \end{figure}

\begin{figure}
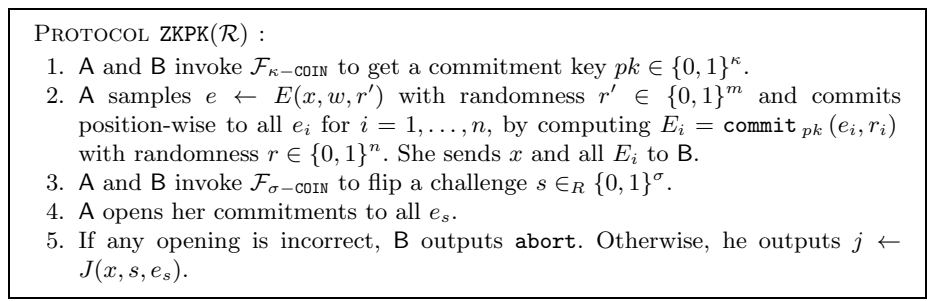

    \small
    \begin{framed}\vspace{-1ex}
      \noindent{\sc Protocol $\ZKPK$ :}\\[-4.5ex]
      \begin{enumerate}
      \item
	$\A$ and $\B$ invoke $\cxF{\kappa}$ to get a commitment key
	$pk \in \zo^\kappa$.
      \item
	$\A$ samples $e \leftarrow E(x,w,r')$ with randomness $r' \in
	\zo^m$ and commits position-wise to all $e_i$ for $i =
	1,\ldots,n$, by computing $E_i =
	\commitk{e_i}{r_i}{pk}$ with randomness $r \in \zo^n$. She
	sends $x$ and all $E_i$ to $\B$.
      \item
	$\A$ and $\B$ invoke $\cxF{\sigma}$ to flip a challenge $s
	\in_R \zo^\sigma$.
      \item
	$\A$ opens her commitments to all $e_s$.
      \item
	If any opening is incorrect, $\B$ outputs $\abort$. Otherwise, he
	outputs $j \la J(x,s,e_s)$.
      \end{enumerate}
      \vspace{-3ex}
    \end{framed}
    \vspace{-4ex}
    \small
    \caption{Zero-Knowledge Proof of Knowledge.}
    \label{fig:zkpk}
  \end{figure}

  Protocol $\ZKPK$ is describe in Fig.\ \ref{fig:zkpk}. It is based on
  our fully simulatable coin-flip protocol $\pi^{(\force,\force)}$,
  which we analyze here in the hybrid model by invoking the ideal
  functionality of sequential coin-flipping twice (but with different
  output lengths).\footnote{Note that in the hybrid model, a simulator
    can enforce a particular outcome to hit also when invoking the
    ideal coin-flip functionality. We then use
    Definition~\ref{def:force} to replace the ideal functionality by
    the actual protocol $\pi^{(\force,\force)}$.} One call to the
  ideal functionality $\cxF{\kappa}$ with output length $\kappa$ is
  required to instantiate a mixed bit commitment scheme
  $\Commit_{pk}$. The second call to the functionality $\cxF{\sigma}$
  produces $\sigma$-bit challenges for a simulatable witness encoding
  scheme with $(E,D,S,J,\hat{E})$ as specified in the previous
  Section~\ref{sec:zkpk.witness.encoding}. The formal proof of
  Theorem~\ref{thm:zkpk} can be found in Appendix~\ref{app:proof.zkpk}. Corollary~\ref{cor:zkpk} follows immediately.
  \begin{theorem}
    \label{thm:zkpk}
    For any simulatable witness encoding scheme $(E,D,S,J,\hat{E})$,
    satisfying completeness, extractability, and simulatability
    according to Definitions~\ref{def:zkpk.complete}
    -~\ref{def:simulate}, and for negligible knowledge error
    $2^{-\sigma}$, protocol $\ZKPK$ securely implements $\zkpkF$.
  \end{theorem}

  \begin{corollary}
    \label{cor:zkpk}
    If there exist mixed commitment schemes, then we can construct a
    classical zero-knowledge proof of knowledge against any quantum
    adversary $\dP \in \dPlayerPoly$ without any set-up assumptions.
  \end{corollary}

\section{Application: Two-Party Function Evaluation}
\label{sec:coin.sfe}

Here, we first show that mixed commitments imply a passively secure
oblivious transfer protocol. From such a protocol it is
straightforward to construct a protocol for \emph{any} classical
poly-time function with security against passive quantum
adversaries~\cite{Kilian88}. We then propose a quantum-secure
implementation for evaluating any such function with security against
active quantum adversaries.


\subsection{Oblivious Transfer}
\label{sec:sfe.ot}

  In an oblivious transfer protocol (OT), the sender $\A$ sends two
  messages $m_0$ and $m_1$ to the selector $\B$. $\B$ can choose which
  message to receive, i.e.\ $m_c$ according to his choice bit
  $c$. $\B$ does not learn anything about the other message $m_{1-c}$,
  and $\A$ does not learn $\B$'s choice bit $c$ (see
  Fig.\ \ref{fig:ot}). The protocol is correct, as $\B$ knows $sk_c$
  and $\xtr{C_c}{sk_c} = \xtr{\commitk{m_c}{r_c}{pk_c}}{sk_c} =
  m_c$. Furthermore, it hides the other message $m_{1-c}$ as
  $\commit_{pk_{1-c}}$ is unconditionally hiding for random
  $pk_{1-c}$, except with negligible probability. Last, the choice bit
  is hidden in the sense of quantum-computational indistinguishability
  between keys for the outer commitments, namely a key produced by
  $\GB$ and a random key by $\GH$.

  \begin{figure}
    \small
    \begin{framed}\vspace{-1ex}
      \noindent{\sc Protocol $\OT$ :}\\[-4.5ex]
      \begin{enumerate}
      \item
	$\B$ samples two keys $pk_0$ and $pk_1$ according to his
	choice bit $c$, i.e.\ he samples $pk_c$ as $(pk_c,sk_c) \la
	\GB$ and $pk_{1-c}$ as  $p_{1-c} \la \GH$. He
	sends $(pk_0,pk_1)$ to $\A$.
      \item
	$\A$ commits to her messages $(m_0,m_1)$ by computing
	$C_0 = \commitk{m_0}{r_0}{pk_0}$ and $C_1 =
	\commitk{m_1}{r_1}{pk_1}$. She sends $(C_0,C_1)$ to $\B$.
      \item
	$\B$ computes $\xtr{C_c}{sk_c}$.
      \end{enumerate}
      \vspace{-3ex}
    \end{framed}
    \vspace{-4ex}
    \small
    \caption{Oblivious Transfer based on Mixed Commitments.}
    \label{fig:ot}
  \end{figure}


\subsection{The Protocol}
\label{sec:sfe.protocol}

  Based on protocol $\OT$, we can construct a passively secure
  protocol for any classical poly-time function $f$. Let
  $\Pi^f_{\A,\B}(x_1,r_1,x_2,r_2)$ denote such a protocol between
  parties $\A$ and $\B$ with inputs $x_1$ and $x_2$ and random strings
  $r_1$ and $r_2$, respectively. We show an implementation of the
  ideal functionality $\sfeF$ evaluating---with security against
  active quantum adversaries---any classical poly-time function $f$
  for which there exists a classical passively secure protocol as
  described above. Functionality $\sfeF$ is shown in
  Fig.\ \ref{fig:sfeF}.\footnote{Note that $y$ does not need to be
    kept secure against external observers and also allows the
    adversary to abort depending on the value of $y$. We stress that
    it is no restriction that we consider common outputs nor that we
    leak $y$ to observers. If we want to compute function $g(x_1,x_2)
    = (y_1,y_2)$ where \emph{only} $\A$ ($\B$) learns $y_1$ ($y_2$),
    we evaluate the common output function $y =
    f((x_1,p_1),(x_2,p_2))$ as follows. Public $y$ contains $y_1
    \oplus p_1$ and $y_2 \oplus p_2$, where $p_1$ and $p_2$ are $\A$'s
    and $\B$'s uniformly random additional input of the same length as
    $y_1$ and $y_2$. Thus, the common outputs are one-time pad
    encrypted using pads known only to the party who is to learn the
    result.} The implementation $\Pi^{\SFE(f)}_{\A,\B}$ of $\sfeF$ is
  shown in Fig.\ \ref{fig:sfe}. Corollary~\ref{cor:sfe} is proven in
  Appendix~\ref{app:proof.sfe}.
  \begin{corollary}
    \label{cor:sfe}
    If there exist mixed commitment schemes, then there exists a
    classical implementation of $\sfeF$ for all classical poly-time
    functions $f$ secure, according to
    Definitions~\ref{def:correctness} and \ref{def:comp.security}.
  \end{corollary}

  \begin{figure}
    \small
    \begin{framed}\vspace{-1ex}
      \noindent{\sc Functionality} $\sfeF$ {\sc with honest
	players: }\\
	On input $x_1$ from Alice and $x_2$ from Bob, $\sfeF$ outputs
	$y = f(x_1,x_2)$ to Alice and Bob.\\\\
      \noindent{\sc Functionality} $\sfeF$ {\sc with
      dishonest Alice:}\\[-4.5ex]
      \begin{enumerate}
      \item
	On input $x_1$ from Alice and $x_2$ from Bob, $\sfeF$ outputs
	$y = f(x_1,x_2)$ to Alice.
      \item
	It then waits to receive her second input $\top$ or $\bot$ and
	outputs $y$ or $\bot$ to Bob, respectively.
      \end{enumerate}
      \vspace{-3ex}
    \end{framed}
    \vspace{-4ex}
    \small
    \caption{The Ideal Functionality for Secure Function Evaluation.}
    \label{fig:sfeF}
   \end{figure}

  \begin{figure}
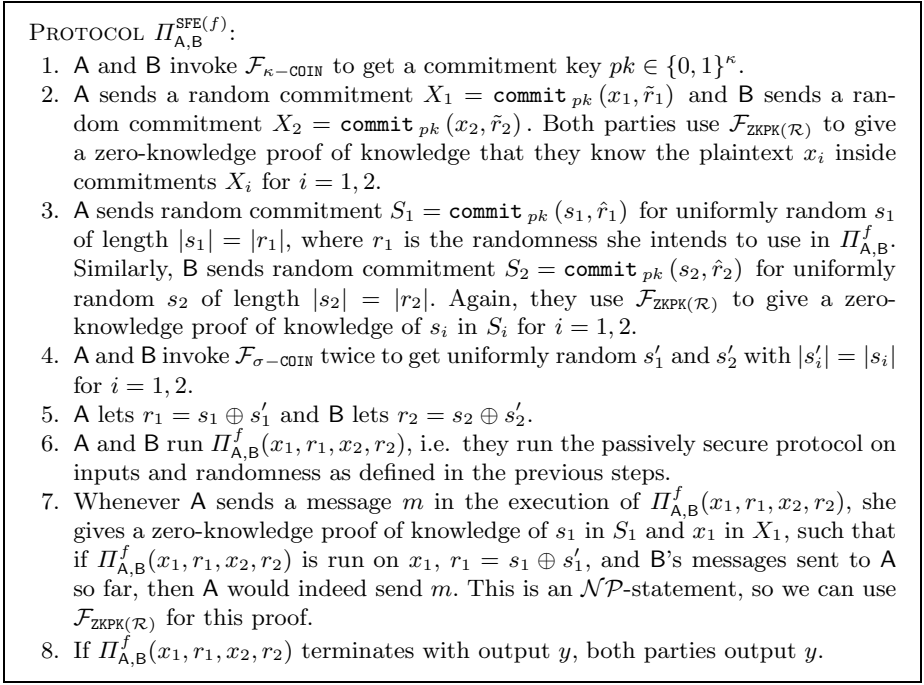

    \small
    \begin{framed}\vspace{-1ex}
      \noindent{\sc Protocol} $\Pi^{\SFE(f)}_{\A,\B}$:
      \\[-4.5ex]
      \begin{enumerate}
      \item
	$\A$ and $\B$ invoke $\cxF{\kappa}$ to get a commitment key
	$pk \in \zo^\kappa$.
      \item
	$\A$ sends a random commitment $X_1 =
	\commitk{x_1}{\tilde{r}_1}{pk}$ and $\B$ sends a random
	commitment $X_2 = \commitk{x_2}{\tilde{r}_2}{pk}$. Both
	parties use $\zkpkF$ to give a zero-knowledge proof of
	knowledge that they know the plaintext $x_i$ inside
	commitments $X_i$ for $i = 1,2$.
      \item
	$\A$ sends random commitment $S_1 =
	\commitk{s_1}{\hat{r}_1}{pk}$ for uniformly random $s_1$ of
	length $\vert s_1 \vert = \vert r_1 \vert$, where $r_1$ is the
	randomness she intends to use in
	$\Pi^f_{\A,\B}$. Similarly, $\B$ sends random
	commitment $S_2 = \commitk{s_2}{\hat{r}_2}{pk}$ for uniformly
	random $s_2$ of length $|s_2| = |r_2|$. Again, they use
	$\zkpkF$ to give a zero-knowledge proof of knowledge of $s_i$
	in $S_i$ for $i = 1,2$.
      \item
	$\A$ and $\B$ invoke $\cxF{\sigma}$ twice to get uniformly
	random $s_1'$ and $s_2'$ with $\vert s_i' \vert = \vert s_i
	\vert$ for $i = 1,2$.
      \item
	$\A$ lets $r_1 = s_1 \oplus s_1'$ and $\B$ lets $r_2 = s_2
	\oplus s_2'$.
      \item
	$\A$ and $\B$ run $\Pi^f_{\A,\B}(x_1,r_1,x_2,r_2)$, i.e.\,
	they run the passively secure protocol on inputs and
	randomness as defined in the previous steps.
      \item
	Whenever $\A$ sends a message $m$ in the execution of
	$\Pi^f_{\A,\B}(x_1,r_1,x_2,r_2)$, she gives a
	zero-knowledge proof of knowledge of $s_1$ in $S_1$ and $x_1$
	in $X_1$, such that if $\Pi^f_{\A,\B}(x_1,r_1,x_2,r_2)$ is
	run on $x_1$, $r_1 = s_1 \oplus s_1'$, and $\B$'s messages
	sent to $\A$ so far, then $\A$ would indeed send $m$. This is
	an \NP-statement, so we can use $\zkpkF$ for this proof.
      \item
	If $\Pi^f_{\A,\B}(x_1,r_1,x_2,r_2)$ terminates with output $y$,
	both parties output $y$.
      \end{enumerate}
      \vspace{-3ex}
    \end{framed}
    \vspace{-4ex}
    \small
    \caption{Procedure for Secure Function Evaluation}
    \label{fig:sfe}
  \end{figure}

\section*{Acknowledgement}

Lunemann acknowledges financial support for part of this work by
Institut Mittag-Leffler, The Royal Swedish Academy of Sciences. Nielsen
acknowledges support from the Danish National Research Foundation and
the National Science Foundation of China (under the grant 61061130540)
for the Sino-Danish Center for the Theory of Interactive Computation,
within which part of this work was performed.


\bibliographystyle{plain}
\bibliography{crypto,qip,procs,personal}


\newpage
\appendix


\section{Proof of Theorem~\ref{thm:force.random}
  (Enforceability and Randomness)}
\label{app:proof.force.random}

\begin{proofx}[ (\emphbf{Enforceability against Alice})]
  In case of corrupted $\dA$, $\dhA$ samples $(pk,sk) \la \GB$ as
  input. It then requests a uniformly random value $h$ from $\clF$. It
  runs $\pi^{(\force,\uncont)}$ with $\dA$, in which $\dhA$ enforces
  the outcome $pk$ in the first step. When $\dA$ sends commitment $A$,
  $\dhA$ uses $sk$ to decrypt $A$ to learn the unique string $a$ that
  $A$ can be opened to. $\dhA$ computes $b = h \oplus a$ and sends $b$
  to $\dA$. If $\dA$ opens commitment $A$ correctly, then the result
  is $c = a \oplus b = a \oplus (h \oplus a) = h$ as desired. In case
  she does not open correctly, $\dhA$ aborts with result
  $\bot$. Otherwise, $\dhA$ outputs whatever $\dA$ outputs.
 
  Since $h$ is uniformly random and independent of $A$ and $a$, it
  follows that $b = h \oplus a$ is uniformly random and independent of
  $A$, exactly as in the protocol. Therefore, the transcript of the
  simulation has the same distribution as the real protocol, except
  that $pk$ is uniform in $\X$ and not in $\zo^\kappa$. This is,
  however, quantum-computationally indistinguishable, as otherwise,
  $\dA$ could distinguish random access to samples from $\X$ from
  random access to samples from $\zo^\kappa$. The formal proof
  proceeds through a series of hybrids as described in full detail in
  the proof for Theorem~\ref{thm:force.force} in
  Appendix~\ref{app:proof.force.force}.

  The above two facts, that first we hit $h$ when we do not abort, and
  second that the transcript of the simulation is
  quantum-computationally indistinguishable from the real protocol,
  show that the resulting protocol is enforceable against Alice and
  simulatable on Alice's side for functionality $\clF$, according to
  Definition~\ref{def:force} combined with Theorem~\ref{def:force}.
\end{proofx}

\begin{proofx}[ (\emphbf{Randomness against Bob})]
  For any $\dB$, $pk$ is uncontrollable, i.e.~$pk \in \zo^\kappa
  \setminus \X$, except with negligible probability, as $\X$ is
  negligible in $\zo^\kappa$. This, in particular, means that the
  commitment $A$ is perfectly hiding the value $a$. Therefore, $a$ is
  uniformly random and independent of $b$, and thus, $h = a \oplus b$
  is uniformly random. This proves that the resulting coin-flip is
  random against Bob, according to Definition~\ref{def:random}.
\end{proofx}


\section{Proof of Theorem~\ref{thm:force.force} (Enforceability)}
\label{app:proof.force.force}

\begin{proofx}[ (\emphbf{Enforceability against Alice})]
  If $\dA$ is corrupted, $\dhA$ samples $(pk,sk) \leftarrow \GB$ as
  input and enforces $\pi^{(\force,\random)}$ in the first step to hit
  the outcome $pk$. It then requests value $h$ from $\clF$. When $\dA$
  sends commitments $(A_1,\ldots,A_\Sigma)$, $\dhA$ uses $sk$ to
  extract $a'$ with $\big( a'_1,\ldots,a'_\Sigma \big) = \big(
  \xtr{A_1}{sk},\ldots,\xtr{A_\Sigma}{sk} \big)$. $\dhA$ then sets $b
  = h \oplus a'$, and sends $b$ to $\dA$. Then $\dhA$ finishes the
  protocol honestly. In the following, we will prove that the
  transcript is quantum-computationally indistinguishable from the
  real protocol and that if $c \neq \bot$, then $c = h$, except with
  negligible probability.

  First, we show indistinguishability. The proof proceeds via a
  hybrid\index{hybrid argument} argument.\footnote{Briefly, a hybrid
  argument is a proof technique to show that two (extreme)
  distributions are computationally indistinguishable via proceeding
  through several (adjacent) hybrid distributions. If all adjacent
  distributions are pairwise computationally indistinguishability, it
  follows by transitivity that the two end points are so as well. We
  want to point out that we are not subject to any restrictions in how
  to obtain the hybrid distributions as long as we maintain
  indistinguishability.} Let $\D{0}$ denote the distribution of the
  output of the simulation as described. We now change the simulation
  such that, instead of sending $b = h \oplus a'$, we simply choose a
  uniformly random $b \in \zo^\ell$ and then output the corresponding
  $h = a' \oplus b$. Let $\D{1}$ denote the distribution of the output
  of the simulation after this change. Since $h$ is uniformly random
  and independent of $a'$ in the first case, it follows that then $b =
  h \oplus a'$ is uniformly random. Therefore, the change to choose a
  uniformly random $b$ in the second case actually does not change the
  distribution at all, and it follows that $\D{0} = \D{1}$.

  By sending a uniformly random $b$, we are in a situation where we do
  not need the decryption key $sk$ to produce $\D{1}$, as we no longer
  need to know $a'$. So we can now make the further change that,
  instead of forcing $\pi^{(\force,\random)}$ to produce a random
  public key $pk \in \X$, we force it to hit a random public key $pk
  \in \zo^\kappa$. This produces a distribution $\D{2}$ of the output
  of the simulation. Since $\D{1}$ and $\D{2}$ only differ in the key
  we enforce $\pi^{(\force,\random)}$ to hit and the simulation is
  quantum poly-time, there exists a poly-sized circuit $Q$, such that
  $Q(\U(\X)) = \D{1}$ and $Q(\U(\zo^\kappa)) = \D{2}$, where $\U(\X)$
  and $\U(\zo^\kappa)$ denote the uniform distribution on $\X$ and the
  uniform distribution on $\zo^\kappa$, respectively. As $\U(\X)$ and
  $\U(\zo^\kappa)$ are quantum-computationally indistinguishable, and
  $Q$ is poly-sized, it follows that $Q(\U(\X))$ and
  $Q(\U(\zo^\kappa))$ are quantum-computationally indistinguishable,
  and therewith, $\D{1} \approxq \D{2}$.

  A last change to the simulation is applied by running
  $\pi^{(\force,\random)}$ honestly instead of enforcing a uniformly
  random $pk \in \zo^\kappa$. Let $\D{3}$ denote the distribution
  obtained after this change. As given in Definition~\ref{def:force},
  real runs of $\pi^{(\force,\random)}$ and runs enforcing a uniformly
  random value are quantum-computationally indistinguishable. Using a
  similar argument as above, where $Q$ is the part of the protocol
  following the run of $\pi^{(\force,\random)}$, we get that $\D{2}
  \approxq \D{3}$. Finally by transitivity, it follows that $\D{0}
  \approxq \D{3}$. The observation that $\D{0}$ is the distribution of
  the simulation and $\D{3}$ is the actual distribution of the real
  protocol concludes the first part of the proof.

  We now argue the second part, i.e., if $c \neq \bot$, then $c = h$,
  except with negligible probability. This follows from
  extractability of the commitment scheme $\Commit_{pk}$.
  Recall that, if $pk \in \X$, then
  the probability that $\dA$ can open any $A$ to a plaintext different
  from $\xtr{A}{sk}$ is at most $(\frac34)^\sigma$ when $S$ is picked
  uniformly at random and independent of $A$. The requirement on $S$
  is however guaranteed (except with negligible probability) by the
  $\random$ flavor of the underlying protocol $\pi^{(\random,\force)}$
  producing $S$. This concludes the proof of enforceability against
  Alice, as given in Definition~\ref{def:force}.
\end{proofx}

\begin{proofx}[ (\emphbf{Enforceability against Bob})]
  To prove enforceability against corrupted $\dB$, we construct a
  simulator $\dhB$ as shown in Fig.\ \ref{fig:force.force.dhB}. It is
  straightforward to verify that the simulation always ensures that $c
  = h$, if $\dB$ does not abort. However, we must explicitly argue
  that the simulation is quantum-computationally indistinguishable
  from the real protocol.

  \begin{figure}
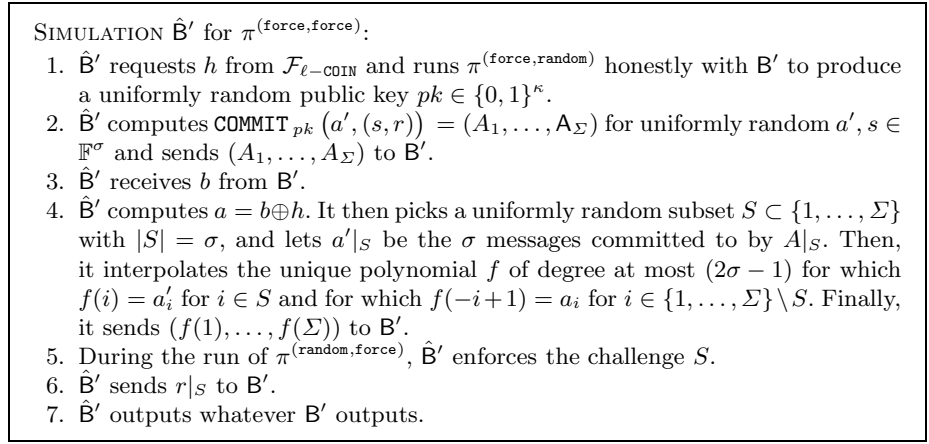

  	\small
    \begin{framed}\vspace{-1ex}
      \noindent{\sc Simulation} $\dhB$ for
      $\pi^{(\force,\force)}$: \\[-4.5ex]
      \begin{enumerate}
      \item
	$\dhB$ requests $h$ from $\clF$ and runs
	$\pi^{(\force,\random)}$ honestly with $\dB$ to produce a
	uniformly random public key $pk \in \zo^\kappa$.
      \item
	$\dhB$ computes $\Commitk{a'}{(s,r)}{pk} =
	(A_1,\ldots,\A_\Sigma)$ for uniformly random $a',s \in
	\FF^\sigma$ and sends $(A_1,\ldots,A_\Sigma)$ to $\dB$.
      \item
	$\dhB$ receives $b$ from $\dB$.
      \item\label{step:trapdoor} $\dhB$ computes $a = b \oplus h$. It
	then picks a uniformly random subset $S \subset
	\set{1,\ldots,\Sigma}$ with $|S| = \sigma$, and lets $a'|_S$
	be the $\sigma$ messages committed to by $A|_S$. Then, it
	interpolates the unique polynomial $f$ of degree at most
	$(2\sigma-1)$ for which $f(i) = a'_i$ for $i \in S$ and for
	which $f(-i+1) = a_i$ for $i \in \set{1,\ldots,\Sigma}
	\setminus S$. Finally, it sends $(f(1), \ldots, f(\Sigma))$ to
	$\dB$.
      \item
	During the run of $\pi^{(\random,\force)}$, $\dhB$ enforces the
	challenge $S$.
      \item\label{step:test.on.S}
	$\dhB$ sends $r|_S$ to $\dB$.
      \item
	$\dhB$ outputs whatever $\dB$ outputs.
      \end{enumerate}
      \vspace{-3ex}
    \end{framed}
    \vspace{-4ex}
    \small
    \caption{Simulation for Bob's $\force$ in $\pi^{(\force,\force)}$.}
    \label{fig:force.force.dhB}
  \end{figure}
 
  Indistinguishability follows by first arguing that the probability
  for $pk \notin \zo^\kappa \setminus \X$ is negligible. This follows
  from $\X$ being negligible in $\zo^\kappa$ and $pk$ produced with
  flavor $\random$ against $\dB$ by $\pi^{(\force,\random)}$ being
  uniformly random in $\zo^\kappa$, except with negligible
  probability.

  Second, we have to show that if $pk \in \zo^\kappa \setminus \X$,
  then the simulation is quantum-computationally close to the real
  protocol. This can be shown via the following hybrid argument. Let
  $\D{0}$ be the distribution of the output of the simulation and let
  $\D{1}$ be the distribution of the output of the simulation where we
  send all $a'_i$ for all $i = \set{1,\ldots,\Sigma}$ at the end of
  Step~(\ref{step:trapdoor}.). Since commitments by $\commit_{pk}$
  are unconditionally hiding in case of $pk\in \zo^\kappa \setminus
  \X$, commitments by $\Commit_{pk}$ are unconditionally
  hiding as well. Furthermore, both $a'$ and $a$ are uniformly
  random, so we obtain statistical closeness between
  $(a',\Commitk{a'}{(s,r)}{pk})$ and
  $(a,\Commitk{a'}{(s,r)}{pk})$. Note further that distributions
  $\D{0}$ and $\D{1}$ can be produced by a poly-sized circuit applied
  to either $(a',\Commitk{a'}{(s,r)}{pk})$ or
  $(a,\Commitk{a'}{(s,r)}{pk}$, it holds that $\D{0} \approxq \D{1}$.

  Now, let $\D{2}$ be the distribution obtained by not simulating the
  opening via the trapdoor, but instead doing it honestly to the value
  committed to, i.e.~$(a',r)$. We still use the challenge $S$ from the
  forced run of $\pi^{(\random,\force)}$ though. However, for
  uniformly random challenges, real runs are quantum-computationally
  indistinguishable from simulated runs, and we get $\D{1} \approxq
  \D{2}$.

  Next, let $\D{3}$ be the distribution of the output of the
  simulation where we run $\pi^{(\random,\force)}$ honestly instead of
  enforcing outcome $S$. We then use the honestly produced $S'$ in the
  proof in Step~(\ref{step:test.on.S}.)~instead of the enforced
  $S$. We can do this, as we modified the process leading to $\D{2}$
  towards an honest opening without any trapdoor, so we no longer need
  to enforce a particular challenge. Under the assumption that
  $\pi^{(\random,\force)}$ is enforceable against $\dB$, and observing
  that real runs are quantum-computationally indistinguishable from
  runs enforcing uniformly random outcomes, we obtain $\D{2} \approxq
  \D{3}$.
 
  It follows by transitivity $\D{0} \approxq \D{3}$, and we
  conclude the proof by observing that after our changes, the process
  producing $\D{3}$ is the real protocol. This concludes the proof of
  enforceability against Bob, according to Definition~\ref{def:force}
  with switched sides.
\end{proofx}


\section{Proof of Theorem~\ref{thm:zkpk} (Zero-Knowledge Proof of Knowledge)}
\label{app:proof.zkpk}

Completeness is obvious. A honest party $\A$, following the protocol
with $(x,w) \in \Rel$ and any valid encoding $e$, will be able to open
all commitments in the positions specified by any challenge
$s$. Honest Bob then outputs $J(x,s,e_s) = \success$.\\

\begin{proofx}[ (\emphbf{Security against dishonest Alice})]
  To prove security in case of corrupted $\dA$, we construct a
  simulator $\dhA$ that simulates a run of the actual protocol with
  $\dA$ and $\zkpkF$. The proof is then twofold. First, we show
  indistinguishability between the distributions of simulation and
  protocol. And second, we verify that the extractability property of
  the underlying witness encoding scheme (see
  Definition~\ref{def:extract}) implies a negligible knowledge
  error. Note that if $\dA$ sends $\abort$ at any point during the
  protocol, $\dhA$ sends some input $(x',w') \notin \Rel$ to $\zkpkF$
  to obtain output $(x,j)$ with $j = \abort$, and the simulation
  halts. Otherwise, the simulation proceeds as shown in
  Fig.\ \ref{fig:simulation.zkpk.dhA}.

  \begin{figure}
  	\small
    \begin{framed}\vspace{-1ex}
      \noindent{\sc Simulation $\dhA$ for
      $\ZKPK$ :}\\[-4.5ex]
      \begin{enumerate}
      \item
	$\dhA$ samples a random key $pk$ along with the extraction key
	$sk$. Then it enforces $pk$ as output from $\cxF{\kappa}$
      \item
	When $\dhA$ receives $x$ and $(E_1,\ldots,E_n)$ from $\dA$, it
	extracts $e = (\xtr{E_1}{sk},\ldots,\xtr{E_n}{sk})$.
      \item
	$\dhA$ completes the simulation by following the protocol
	honestly. If any opening of $\dA$ is incorrect, $\dhA$
	aborts. Otherwise, $\dhA$ inputs $\big( x,D(x,e) \big)$ to
	$\zkpkF$ and receives $(x,j)$ back. $\dhA$ outputs the final
	state of $\dA$ as output in the simulation.
      \end{enumerate}
      \vspace{-3ex}
    \end{framed}
    \vspace{-4ex}
    \small
    \caption{Simulation against dishonest Alice.}
    \label{fig:simulation.zkpk.dhA}
  \end{figure}

  Note that the only difference between the real protocol and the
  simulation is that $\dhA$ uses a random public key $pk$ sampled
  along with an extraction key $sk$, instead of a uniformly random $pk
  \in \zo^\kappa$. It then enforces $\cxF{\kappa}$ to hit
  $pk$. However, by assumption on the commitment keys and by the
  properties of the ideal coin-flip functionality, the transcripts
  of simulation and protocol remain quantum-computationally
  indistinguishable under these changes.

  Next, we analyze the output in more detail. It is clear that
  whenever honest $\B$ would output $\abort$ in the actual protocol,
  also $\dhA$ aborts, namely, if $\dA$ does deviate in the last steps
  of protocol and simulation, respectively. Furthermore, $\dhA$
  accepts if and only if $(x,D(x,e)) \in \Rel$ or in other words, the
  judgment of the functionality is positive, denoted by $j_\F =
  \success$.

  It is therefore only left to prove that the case of $j_\F = \abort$
  but $j_J = \success$ is negligible, where the later denotes the
  judgment of algorithm $J(x,s,e_s)$ as in the protocol. In that case,
  we have $(x,D(x,e)) \notin \Rel$. This means that $w$ is not
  extractable from $D(x,e)$, which in turn implies that
  $(\xtr{E_1}{sk},\ldots,\xtr{E_n}{sk}) = e$ is not admissible. Thus,
  there are no two distinct challenges $s$ and $s'$, in which $\dA$
  could correctly open her commitment to $e$. It follows by
  contradiction that there exists at most one challenge $s$ which
  $\dA$ can answer. We produce $s \in \zo^\sigma$ uniformly at random,
  from which we obtain an acceptance probability of at most
  $2^{-\sigma}$. Thus, we conclude the proof with negligible knowledge
  error, as desired.
  \end{proofx}

\begin{proofx}[ (\emphbf{Security against dishonest Bob})]
  To prove security in case of corrupted $\dB$, we construct simulator
  $\dhB$ as shown in Fig.\ \ref{fig:simulation.zkpk.dhB}. Our aim is
  to verify that this simulation is quantum-computationally
  indistinguishable from the real protocol. The key aspect will be the
  simulatability guarantee of the underlying witness encoding scheme,
  according to Definition~\ref{def:simulate}.
 
  \begin{figure}
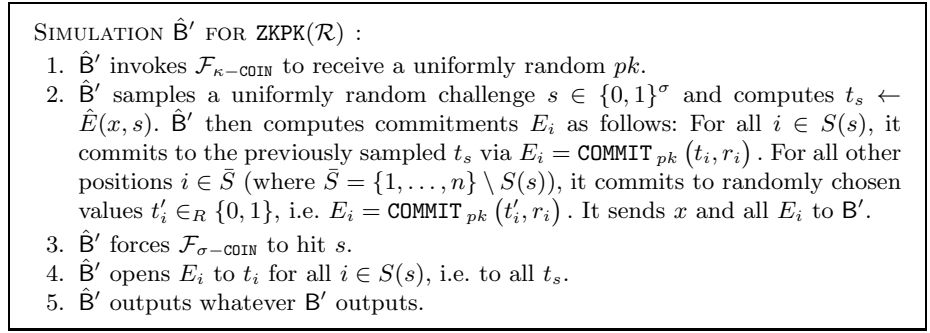

  	\small
    \begin{framed}\vspace{-1ex}
      \noindent{\sc Simulation $\dhB$ for $\ZKPK$
      :}\\[-4.5ex]
      \begin{enumerate}
      \item
	$\dhB$ invokes $\cxF{\kappa}$ to receive a uniformly random
	$pk$.
      \item
	$\dhB$ samples a uniformly random challenge $s \in \zo^\sigma$
	and computes $t_s \la \hat{E}(x,s)$. $\dhB$ then computes
	commitments $E_i$ as follows: For all $i \in S(s)$, it commits
	to the previously sampled $t_s$ via $E_i =
	\Commitk{t_i}{r_i}{pk}$. For all other positions $i \in
	\bar{S}$ (where $\bar{S} = \set{1,\ldots,n} \setminus S(s)$),
	it commits to randomly chosen values $t'_i \in_R \zo$,
	i.e.~$E_i = \Commitk{t'_i}{r_i}{pk}$. It sends $x$ and all
	$E_i$ to $\dB$.
      \item
	$\dhB$ forces $\cxF{\sigma}$ to hit $s$.
      \item
	$\dhB$ opens $E_i$ to $t_i$ for all $i \in S(s)$, i.e.\ to all
	$t_s$.
      \item $\dhB$ outputs whatever $\dB$ outputs.
      \end{enumerate}
      \vspace{-3ex}
    \end{framed}
    \vspace{-4ex}
    \small
    \caption{Simulation against dishonest Bob.}
    \label{fig:simulation.zkpk.dhB}
  \end{figure}

  The proof proceeds via a hybrid argument. Let $\D{0}$ be the
  distribution of the simulation as described in
  Fig.\ \ref{fig:simulation.zkpk.dhB}. Let $\D{1}$ be the
  distribution obtained from the simulation but with the following
  change: We inspect $\zkpkF$ to get a valid witness $w$ for instance
  $x$, and let $e \la E(x,w,r')$ be the corresponding encoding. Note
  that this is possible as a thought experiment for any adjacent
  distribution in a hybrid argument. From $e$ we then use bits $e_s$
  for the same $S(s)$ as previously, instead of bits $t_s$ sampled by
  $\hat{E}(x,s)$. All other steps are simulated as before. By the
  simulatability of the encoding scheme
  (Definition~\ref{def:simulate}), it holds that the bits $t_s$ in
  $\D{0}$ and the bits $e_s$ in $\D{1}$ have the same
  distribution. Thus, we obtain $\D{0} = \D{1}$.

  We further change the simulation in that we compute the bits in all
  positions $i \in \bar{S}$ by $e_i$ of the encoding $e$ defined in
  the previous step. Again, all other steps of the simulation remain
  unchanged. Let $\D{2}$ denote the new distribution. The only
  difference now is that for $i \in \bar{S}$, the commitments $E_i$
  are to the bits $e_i$ of a valid $e$ and not to uniformly random
  bits $t'_i$. This, however, is quantum-computationally
  indistinguishable to $\dB$ for $pk \in_R \zo^\kappa$, as $\Commit$
  is quantum-computationally hiding towards $\dB$. Note that $pk$ is
  guaranteed to be random by an honest call to $\cxF{\kappa}$ and
  recall that we do not have to open the commitments in these
  positions. Hence, we get that $\D{1} \approxq \D{2}$.

  Note that after the two changes, leading to distributions $\D{1}$
  and $\D{2}$, the commitment step and its opening now proceed as in
  the actual protocol, namely, we commit to the bits of $e \la
  E(x,e,r')$ and open the subset corresponding to $S(s)$. The
  remaining difference to the real protocol is the enforcement of
  challenge $s$, whereas $s$ is chosen randomly in the protocol. Now,
  let $\D{3}$ be the distribution of the modified simulation, in which
  we implement this additional change of invoking $\cxF{\sigma}$
  honestly and then open honestly to the resulting $s$. Note that both
  processes, i.e., first choosing a random $s$ and then enforcing it
  from $\cxF{\sigma}$, or invoking $\cxF{\sigma}$ honestly and
  receiving a random $s$, result in a uniformly random distribution on
  the output of $\cxF{\sigma}$. Thus, we obtain $\D{2} = \D{3}$.

  By transitivity, we conclude that $\D{0} \approxq \D{3}$, and
  therewith, that the simulation is quantum-computationally
  indistinguishable from the actual protocol.
\end{proofx}


\section{Proof of Corollary~\ref{cor:sfe} (Two-Party Function Evaluation)}
\label{app:proof.sfe}

\begin{proofx}[ (\emphbf{Security against dishonest Alice})]
  If $\dA$ is corrupted, $\dhA$ uses the proof of knowledge to learn
  her $x_1$ inside commitment $X_1$. Then $\dhA$ inputs $x_1$ to
  $\sfeF$ as $\dA$'s input and receives $y = f(x_1,x_2)$. Now, $\dhA$
  invokes $\Sim^f_{\dhA,\hB}$ with input $x_1$ and $y$. This, in
  particular, yields randomness $r_1$ and is quantum-computationally
  indistinguishable from a real run of protocol
  $\Pi^f_{\dA,\B}$. Furthermore, the simulated transcript contains all
  messages sent by $\hB$. Next, $\dhA$ uses the proof of knowledge to
  learn $\dA$'s $s_1$ inside commitment $S_1$. Then $\dhA$ enforces
  challenge $s_1'$ such that $s_1' = s_1 \oplus r_1$, and thereby
  forces $\dA$ to use $r_1$ in the following.
 
  $\dhA$ now runs $\Pi^f_{\dA,\B}$ with $\dA$. Whenever it is the turn
  of $\hB$ to send a message, $\dhA$ sends the next message obtained
  already by $\Sim^f_{\dhA,\hB}$. Whenever it is the turn of $\dA$ to
  send a message $m$, $\dhA$ checks whether it coincides with the
  message obtained already by $\Sim^f_{\dhA,\hB}$. Note that by
  construction her only consistent message really is the message
  obtained previously. In case of inconsistency, $\dA$ will fail in
  her following proof of knowledge, where she must prove that $m$ is
  consistent with $x_1$ in $X_1$, $s_1$ in $S_1$, and where $r_1 = s_1
  \oplus s_1'$ with $r_1$ obtained from $\Sim^f_{\dhA,\hB}$. Hence, if
  $\dA$ does not send an inconsistent $m$ and thereby make the
  protocol fail, then the transcript of this simulation is consistent
  with the previous invocation of $\Sim^f_{\dhA,\hB}$. In that case,
  $\dhA$ inputs $\top$ as second input to $\sfeF$, which outputs $y$
  as final result. Otherwise, the input is $\bot$, yielding output
  $\bot$ from $\sfeF$ and modeling the case where a wrong $m$ makes
  $\dA$ fail in the proof of knowledge.
 
  Therefore, the only difference between the simulation with $\sfeF$ and
  the real procedure $\Pi_{\dA,\B}^{\SFE(f)}$ is $\dA$'s views,
  simulated by $\Sim^f_{\dhA,\hB}$ and actually produced by
  $\Pi^f_{\dA,\B}$, respectively. These views, however, are by
  assumption quantum-computationally indistinguishable.
\end{proofx}


\end{document}